\documentclass[lettersize,journal]{IEEEtran}
\usepackage{amsmath,amsfonts}
\usepackage{algorithmic}
\usepackage{algorithm}
\usepackage{array}
\usepackage[caption=false,font=normalsize,labelfont=sf,textfont=sf]{subfig}
\usepackage{textcomp}
\usepackage{stfloats}
\usepackage{url}
\usepackage{verbatim}
\usepackage{graphicx}
\usepackage{cite}
\usepackage{multirow}
\usepackage[colorlinks,
            linkcolor=blue,
            anchorcolor=blue,
            citecolor=blue]{hyperref}
\usepackage{float}
\usepackage{bm}

\begin{document}

\title{Geometry-based spherical JND modeling for 360° display}

\author{Hongan~Wei,
	    Jiaqi~Liu,
		Bo~Chen,
	    Liqun~Lin,~\IEEEmembership{Member,~IEEE,}
        Weiling~Chen,~\IEEEmembership{Member,~IEEE,}
	    and~Tiesong Zhao,~\IEEEmembership{Senior Member,~IEEE}
        % <-this % stops a space
\thanks{This work was supported Natural Science Foundation of Fujian Province, China (Grant No. 2022J02015) and in part by National Natural Science Foundation of China (Grant No. 62171134). \it{(\it Corresponding author: Liqun Lin.)}}
% <-this % stops a space
\thanks{H. Wei, J. Liu,  L. Lin, W. Chen and T. Zhao are with the Fujian Key Lab for Intelligent Processing and Wireless Transmission of Media Information, College of Physics and Information Engineering, Fuzhou University, Fuzhou, Fujian 350116, China and also with Fujian Science \& Technology Innovation Laboratory for Optoelectronic Information of China (e-mails: \{weihongan, 201127045, lin\_liqun, t.zhao\}@fzu.edu.cn)\par
B. Chen is with the Department of Computer Science,University of Illinois at Urbana-Champaign, Urbana, IL, USA (email:boc2@illinois.edu)}
}

% The paper headers
\markboth{Journal of \LaTeX\ Class Files,~Vol.~14, No.~8, August~2021}%
{Shell \MakeLowercase{\textit{et al.}}: A Sample Article Using IEEEtran.cls for IEEE Journals}

\IEEEpubid{}
% Remember, if you use this you must call \IEEEpubidadjcol in the second
% column for its text to clear the IEEEpubid mark.

\maketitle

\begin{abstract}
360° videos have received widespread attention due to its realistic and immersive experiences for users. To date, how to accurately model the user perceptions on 360° display is still a challenging issue. In this paper, we exploit the visual characteristics of 360° projection and display and extend the popular just noticeable difference (JND) model to spherical JND (SJND). First, we propose a quantitative 2D-JND model by jointly considering spatial contrast sensitivity, luminance adaptation and texture masking effect. In particular, our model introduces an entropy-based region classification and utilizes different parameters for different types of regions for better modeling performance. Second, we extend our 2D-JND model to SJND by jointly exploiting latitude projection and field of view during 360° display. With this operation, SJND reflects both the characteristics of human vision system and the 360° display. Third, our SJND model is more consistent with user perceptions during subjective test and also shows more tolerance in distortions with fewer bit rates during 360° video compression. To further examine the effectiveness of our SJND model, we embed it in Versatile Video Coding (VVC) compression. Compared with the state-of-the-arts, our SJND-VVC framework significantly reduced the bit rate with negligible loss in visual quality.
 
\end{abstract}


\section{Introduction}
\IEEEPARstart{360°}{video} has received widespread attention from academia and industry for its immersive and interactive experience \cite{ref1, ref2}. Unlike 2D video, the spherical 360° videos are firstly mapped onto 2D plane using a variety of projection formats. Currently, Equirectangular Projection (ERP) is the most widely used in 360° videos. To provide highly immersive experience, a high resolution for the 360° contents is required, which makes  360° videos more bandwidth consumption than 2D-conventional videos\cite{ref3}. In addition,  360° videos are viewed through a viewport, which is completely different from 2D-conventional videos. The above characteristics of 360° videos set new challenges in quality assessment and subjective perception coding of 360° videos.

Currently, there are many attempts to exploit the structural features of 360° videos, such as allocating different bit resources to various latitude regions \cite{ref4, ref5, ref6} and spatially dividing the scene into several separate tiles\cite{ref7, ref8}. Tiles inside users’ FoV are delivered at higher quality level, while those out of sight are transmitted at lower quality level or ignored. However, the above great efforts focus on eliminating spatio-temporal and statistical redundancy without taking into account the perceptual redundancy. In particular, the Just Noticeable Distortion (JND) relies on the mechanism of Human Visual System (HVS) that humans cannot perceive small visual signal variations within certain visual contents. It has been widely applied in perceptual image/video compression\cite{ref9, ref10, ref11,ref12}.
\begin{figure}[t]
	\centering
	\subfloat[]{
		\includegraphics[width=4.1cm, height = 2.0cm]{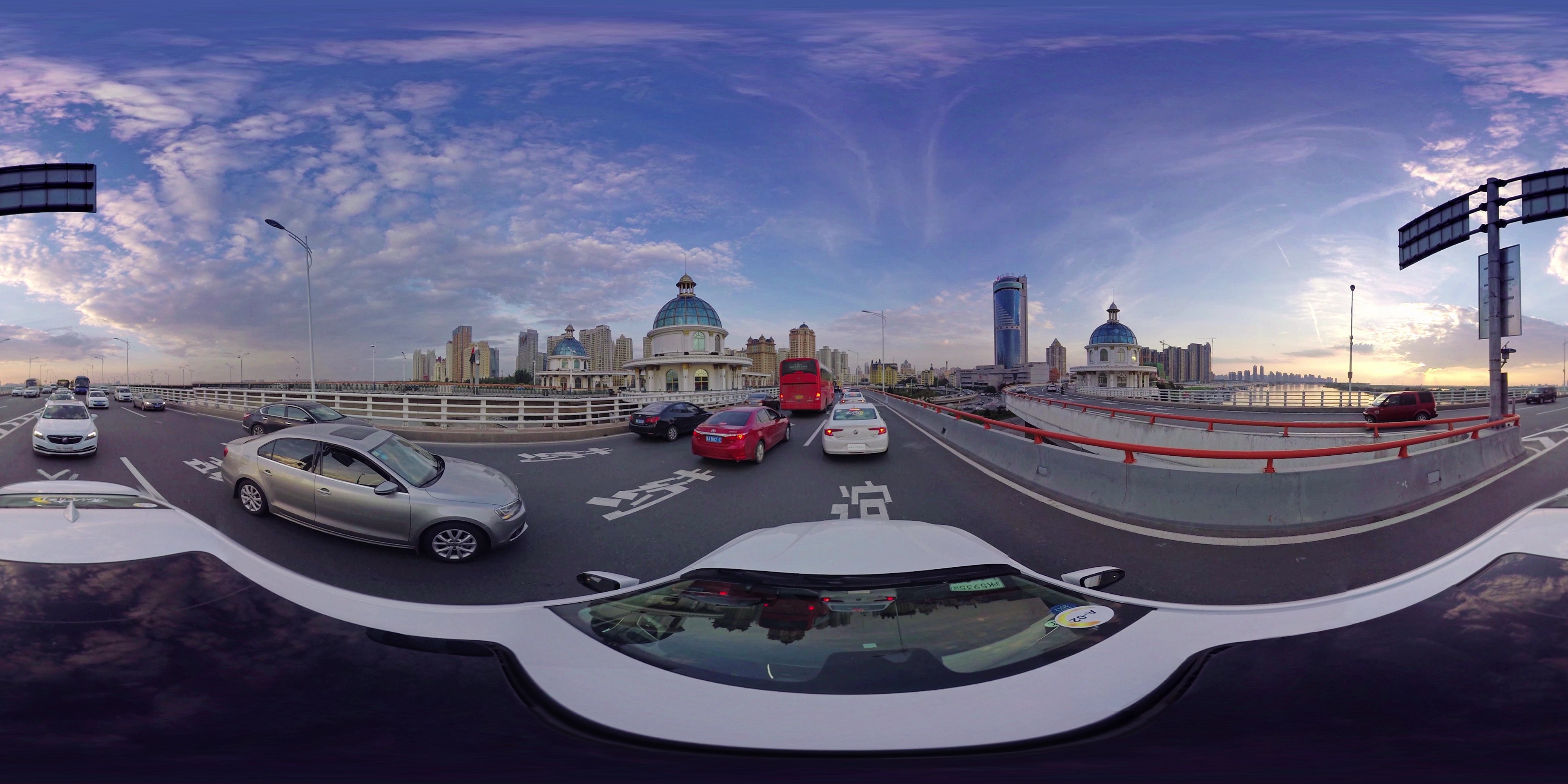}
		
	}
	\subfloat[]{
		\includegraphics[width=4.1cm, height = 2.0cm]{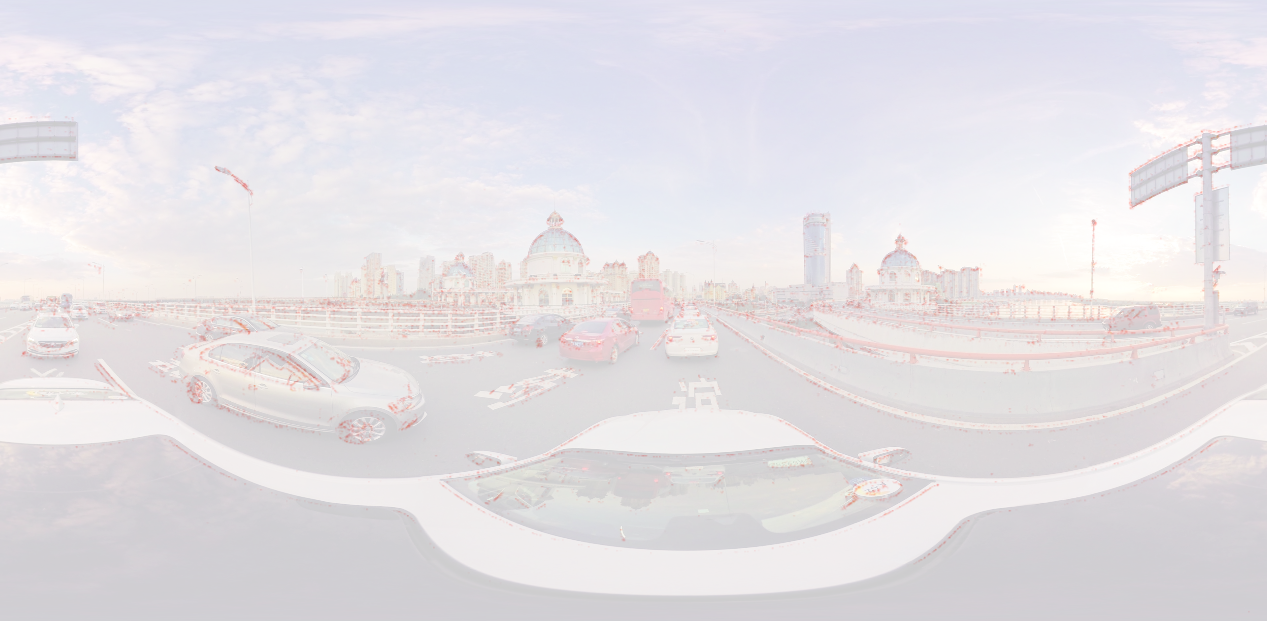}
		
	}
	\quad
	\subfloat[]{		
		\includegraphics[width=4.1cm, height = 2.0cm]{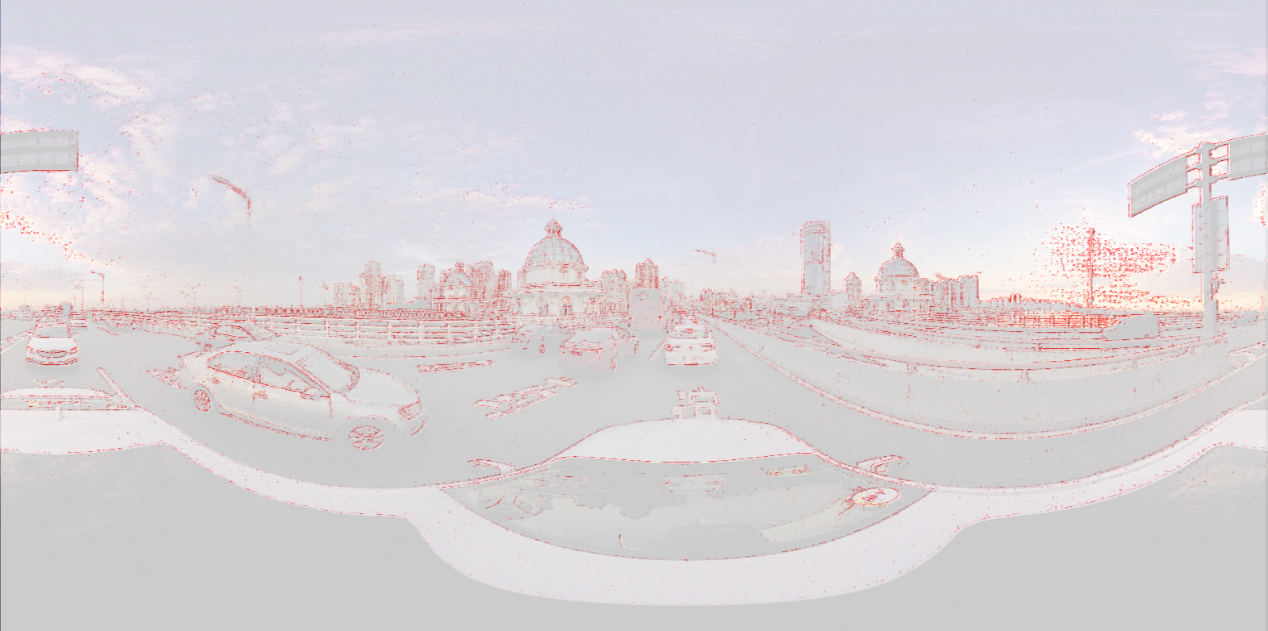}
		
	}
	\subfloat[]{
		\includegraphics[width=4.1cm, height = 2.0cm]{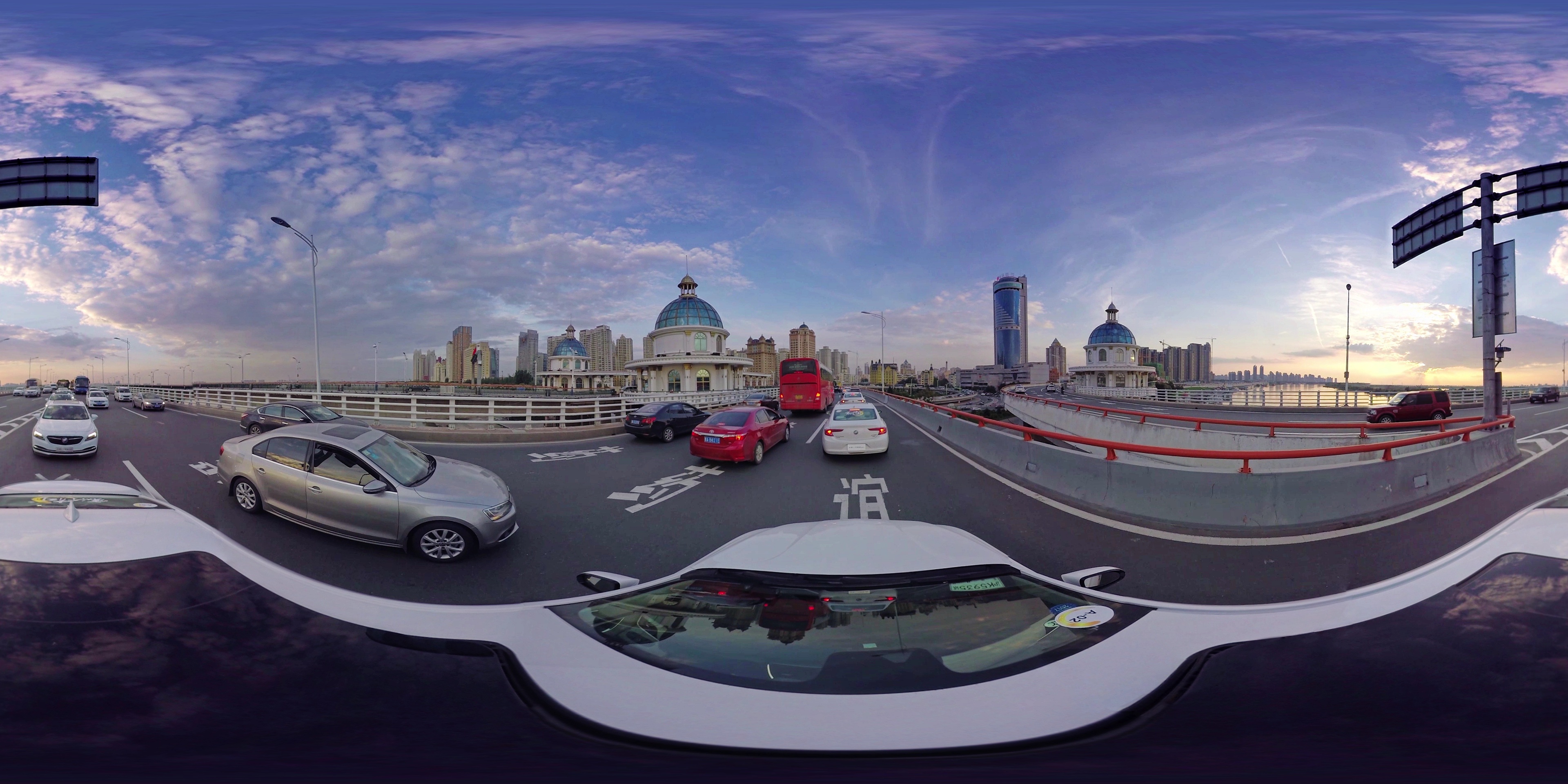}
		
	}
	\caption{Our proposed JND model and its application: (a) an original 360° video; (b) 2D-JND modeling of (a); (c) SJND modeling of (a); (d) Compressed and reconstructed frame with SJND, where its bitrate is reduced by 3.10\% compared with that of original codec.}
	\label{fig1}
\end{figure}

The conventional JND models can be mainly categorized into pixel-wise models and sub-band JND models. Pixel-wise models calculate the JND threshold for each pixel directly. Sub-band JND models calculate the JND threshold for each sub-band coefficient by transforming an image into frequency domain representation. Pixel domain JND models mainly consider the Luminance Adaptation (LA) and the Contrast Masking (CM) effects. An inchoate pixel domain method was proposed in \cite{ref13}, where LA, CM based on canny edge and temporal masking were combined to generate the JND masking. Wu et al.\cite{ref14} developed a disorderly concealment JND model with free-energy principle, which took luminance adaptation and spatial masking into consideration. Reference \cite{ref15} introduced a novel JND model by using the effect from content regularity, luminance adaptation and contrast masking. While such JND models improve estimation accuracy to some extent, but fail to take the Contrast Sensitivity Function (CSF).

In addition, subband-based JND models first transfer an image to a subband domain(i.e. ,  DCT domain), which usually take CSF, LA and CM into consideration and are widely applied to image and video compression. Reference \cite{ref16} proposed a JND estimation model considering multiple masking effects. Wang et al. \cite{ref17} presented an efficient JND estimation approach for screen content images by considering high-frequency sensitivity and orientation sensitivity correction. A novel DCT-based JND profile was build in \cite{ref18}, considering the temporal masking and Foveated Masking (FM). The JND model\cite{ref19} was designed by taking three AC DCT coefficients of the DCT-based CM. These models are suitable for compression of images/videos, but are for 2D images and videos.

In summary, the existing studies on JND models are mainly focused on traditional 2D videos/images. JND measures the visibility threshold under the assumption that visual acuity remains constant over the whole image. 2D-JND modeling only takes into account the differences of each pixel and the role of neighboring pixels, without considering the spatial location of pixels. However, Spherical JND (SJND) model is closely related to the spatial location of the pixels\cite{ref20}. The 2D-JND models cannot be directly used for  360° videos. SJND models still lacks research. To address this issue, we propose a SJND model for 360° display. In addition, we further demonstrate its application in video coding. The flow chart is illustrated in Fig.1. The contributions of this work are summarized as follows: 
\begin{itemize}
	\item The first JND model for 360° display. There have been lots of studies on 2D-JND that which treat all pixel locations evenly. However, the user attentions are critical in the perception of 360° display, which make it imperative to develop an effective JND model for 360° videos.
	\item An analytical mapping from 2D-JND to SJND. The state-of-the-art 360° videos are usually warped into 2D videos for compression. During this process, the latitude features and filed of view are essential factors to influence user perception. Based on these effects, we propose to map a 2D-JND model to SJND.
	\item Extensive experiments to validate our SJND model. We perform both subjective and objective experiments to compare SJND and the state-of-the-art methods, in which SJND shows significantly superior performance. In addition, we also incorporate this model in the most recent VVC encoder, which significantly reduces its bit rate with negligible loss in perceived visual quality.
\end{itemize}

\section{The Proposed SJND Model}
The state-of-the-art 360° videos are usually warped into 2D videos for compression. During this process, the latitude features and filed of view are essential factors to influence user perception. Based on these effects, we propose to map a 2D-JND model to SJND model.

\subsection{Texture Masking Factor}
Existing 2D images/videos JND models are basically based on CSF, LA and TM. However, these models achieve data compression mainly by reducing the hierarchical information representing image luminance (or chroma), which cannot accurately estimate the texture masking effect. The texture masking factor refers to the fact that the HVS exhibits different degrees of sensitivity to distortion in areas of texture of varying complexity. In general, the HVS is more sensitive in plain and contour areas, which has  relatively lower JND thresholds. Thus, the distortion is less. Studies have shown that image information entropy can effectively reflect the complexity and information content of an image\cite{ref23, ref24}. 

Based on the above analysis, we utilize image information entropy to characterize the complexity of image blocks. In this paper, we calculate the entropy value of each DCT block as follows:
\begin{equation}
	e_{i,j} = - \sum_{i}^{N} \sum_{j}^{N} p(i,j) \log p(i,j),
\end{equation}
\begin{equation}
	p(i,j) = \frac {I(i,j)}{\sum\limits_{i}^{N} \sum\limits_{j}^{N} I(i,j)},
\end{equation}
where $ e(i, j) $ denotes the information entropy of a DCT block; $ p(i, j) $ is the probability distribution of gray level at point $ (i, j) $. $ I(i, j) $ is the gray level at point $ (i, j) $ in an image, $ N $ is the dimension of a DCT block.

\begin{table}[htbp]
	\centering
	\caption{The Entropy Values of Image Blocks.}
	\renewcommand{\arraystretch}{2.0}
	\resizebox{82mm}{28mm}{
		\begin{tabular}{lcclcc}
			\hline
			Video Sequence                                             & \multicolumn{1}{l}{Image Patch} & \multicolumn{1}{l}{Entropy Value} & Video Sequence                            & \multicolumn{1}{l}{Image Patch} & \multicolumn{1}{l}{Entropy Value} \\
			\hline
			\multicolumn{1}{c}{\multirow{3}{*}{ArenaOfValor}}  \vspace{0.8mm} & {\raisebox{-.5\height}{\includegraphics[width=0.75cm]{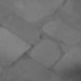}}}       & 2.99                   & \multirow{3}{*}{BasketballDrive} & {\raisebox{-.5\height}{\includegraphics[width=0.75cm]{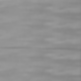}}}          & 2.15                   \\
			\multicolumn{1}{c}{}                             \vspace{0.8mm} & {\raisebox{-.5\height}{\includegraphics[width=0.75cm]{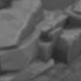}}}       & 3.93                   &                                  & {\raisebox{-.5\height}{\includegraphics[width=0.75cm]{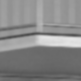}}}          & 3.43                   \\
			\multicolumn{1}{c}{}                              \vspace{0.8mm} & {\raisebox{-.5\height}{\includegraphics[width=0.75cm]{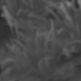}}}       & 4.36                   &                                  & {\raisebox{-.5\height}{\includegraphics[width=0.75cm]{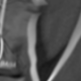}}}          & 4.23                  \\
			\hline
			\multirow{3}{*}{BQTrace}                   \vspace{0.8mm}  & {\raisebox{-.5\height}{\includegraphics[width=0.75cm]{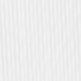}}}       & 1.98                   & \multirow{3}{*}{Cactus}          & {\raisebox{-.5\height}{\includegraphics[width=0.75cm]{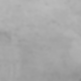}}}         & 1.75                   \\
			\vspace{0.8mm}      & {\raisebox{-.5\height}{\includegraphics[width=0.75cm]{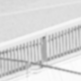}}}       & 3.78                   &                                  & {\raisebox{-.5\height}{\includegraphics[width=0.75cm]{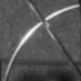}}}         & 3.99                  \\
			\vspace{0.8mm}     & {\raisebox{-.5\height}{\includegraphics[width=0.75cm]{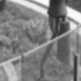}}}       & 4.87                   &                                  & {\raisebox{-.5\height}{\includegraphics[width=0.75cm]{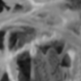}}}         & 4.84                   \\
			
			\hline
			\multirow{3}{*}{MarketPlace}                 \vspace{0.8mm}     & {\raisebox{-.5\height}{\includegraphics[width=0.75cm]{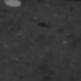}}}      & 3.01                   & \multirow{3}{*}{RitualDance}     & {\raisebox{-.5\height}{\includegraphics[width=0.75cm]{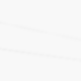}}}         & 0.48                   \\
			\vspace{0.8mm}     & {\raisebox{-.5\height}{\includegraphics[width=0.75cm]{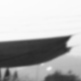}}}      & 2.97                   &                                  & {\raisebox{-.5\height}{\includegraphics[width=0.75cm]{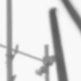}}} 		& 2.87                  \\
			\vspace{0.8mm}     & {\raisebox{-.5\height}{\includegraphics[width=0.75cm]{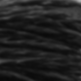}}}      & 4.26                   &                                  & {\raisebox{-.5\height}{\includegraphics[width=0.75cm]{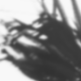}}} 		& 4.34                  \\
			
			\hline
		\end{tabular}
	}
\end{table}

Table 1 shows the information entropy values of Versatile Video Coding (VVC) coding standard sequence videos with different texture complexity. Based on the results, we classify an image into different types. For texture areas, the HVS are less sensitive to the low frequency distortion, such as blocking artifacts, but the high frequency information should be retained. For the intra-band masking effect, we take into account the fact that the high spatial frequency will reduce the contrast masking effect, especially for the high frequency energy areas in a DCT block. The correction factor of the intra-band masking is added in each block. The final contrast masking factor $ \alpha_{\rm TM} $ can be determined:
\begin{equation}
	\rm \alpha_{TM}  = \left\{
	\begin{array}{lll}
		\varepsilon \left.\  ,\mbox{$ (i^{2} + j^{2}) \leq 16 $ }  \mbox{in plain and texture blocks} \right.\\[2mm]
		\varepsilon \cdot min(4, max(1, (\frac {C(k, n, i, j)}{\rm JND_{base} \times \alpha_{\rm {LA}}})^{0.36})), \left.\ \mbox{others} \right. \\
	\end{array} , \right.
\end{equation}
$ C(k,n,i, j) $ is the $ (i, j)$-th DCT coefficient in the $ n $-th block of the $ k $-th frame.

\subsection{ Latitude Characteristic Modeling}
In order to comply with current video encoding standards, we need to project  a 360° video as 2D images for encoding, back-project the 2D images back to the 360° video when viewing. This projection process may result in a change JND threshold, i.e., geometric mapping characteristics. ERP mapping is often used as a storage format for 360° videos due to its low computational complexity and ease of image processing. However, the projection introduces some geometric distortions, as indicated by the red rectangles in Fig.2. In particular, the straight lines in the spherical domain are rendered as curved straight lines. In addition, the sampling densities are higher in the polar regions compared to the equatorial regions. Therefore, it is necessary to investigate the effect on JND thresholds in different latitude regions.
\begin{figure}[h]
	\vspace{-2mm}
	\centering
	\subfloat[Sampling on ERP projection and spherical surface]{
		\includegraphics[width=6.8cm,height=2.8cm]{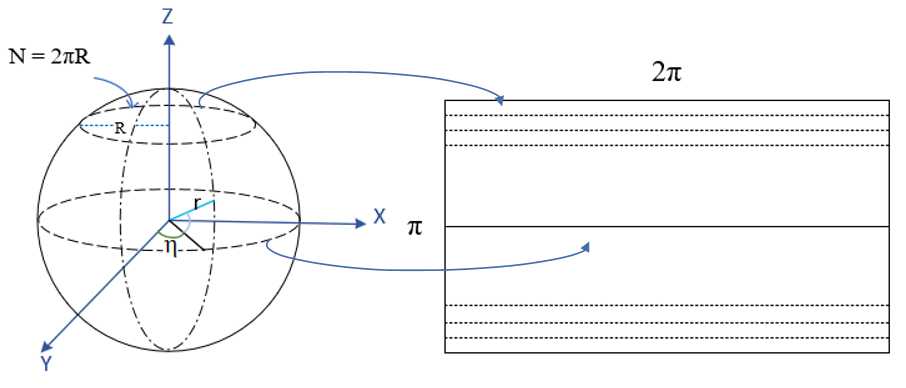}
		\label{c}
	}
	\quad
	\subfloat[ERP image]{
		\includegraphics[width=6.8cm,height=2.8cm]{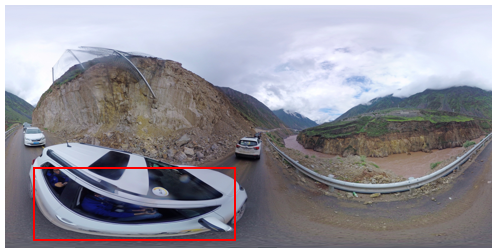}
		\label{a}
	}
	\caption{ERP Projection.}
	\label{fig2}
\end{figure}

{\bfseries Latitude characteristics:} To study the effect of sample density on spherical JND threshold, we design experiments to downsample the original images by 1/2, 1/4, 1/8, 1/16, respectively, and calculate threshold using the proposed 2D-JND model. The experimental results are shown in Table 2. As can be seen from it, when the sampling density of an image is smaller, the JND value is also smaller, that is, the less distortion the image can accommodate; when the sampling density is larger, the JND threshold value is larger. For a 360° images in ERP format, the sampling density near the equator is similar to that on the sphere, while the sampling density in regions far from the equator is much greater than the ideal sampling density on the sphere. Therefore, the JND values in the polar regions increase with latitude, further indicating that the spherical JND threshold is related to the latitude change.
\begin{table}[htpb]
	%\vspace{-6.5mm}
	\centering
	\caption{2D-JND Threshold for The Original and Downsampled Images.}
	\renewcommand{\arraystretch}{1.1}
	\resizebox{82mm}{24mm}{
		\begin{tabular}{lcclcc}
			\hline
			Image                                             & \multicolumn{1}{l}{Downsample} & \multicolumn{1}{l}{2D-JND} & Image                            & \multicolumn{1}{l}{Downsample} & \multicolumn{1}{l}{2D-JND} \\
			\hline
			\multicolumn{1}{c}{\multirow{5}{*}{ArenaOfValor}} & Original                       & 17.51                   & \multirow{5}{*}{BasketballDrive} & Original                       & 16.89                   \\
			\multicolumn{1}{c}{}                              & 1/2                            & 17.10                   &                                  & 1/2                            & 16.09                   \\
			\multicolumn{1}{c}{}                              & 1/4                            & 16.65                   &                                  & 1/4                            & 15.75                   \\
			\multicolumn{1}{c}{}                              & 1/8                            & 16.37                   &                                  & 1/8                            & 15.41                   \\
			\multicolumn{1}{c}{}                              & 1/16                           & 16.14                   &                                  & 1/16                           & 15.34                   \\
			\hline
			\multirow{5}{*}{BQTrace}                          & Original                       & 14.87                   & \multirow{5}{*}{Cactus}          & Original                       & 17.27                   \\
			& 1/2                            & 14.42                   &                                  & 1/2                            & 16.65                   \\
			& 1/4                            & 13.95                   &                                  & 1/4                            & 15.96                   \\
			& 1/8                            & 13.57                   &                                  & 1/8                            & 15.51                   \\
			& 1/16                           & 13.55                   &                                  & 1/16                           & 15.44                   \\
			\hline
			\multirow{5}{*}{MarketPlace}                      & Original                       & 18.45                   & \multirow{5}{*}{RitualDance}     & Original                       & 15.73                   \\
			& 1/2                            & 18.24                   &                                  & 1/2                            & 15.60                   \\
			& 1/4                            & 18.16                   &                                  & 1/4                            & 15.52                   \\
			& 1/8                            & 18.11                   &                                  & 1/8                            & 15.46                   \\
			& 1/16                           & 17.95                   &                                  & 1/16                           & 15.31                   \\
			\hline
		\end{tabular}
	}
\end{table}

In order to further verify the JND variable law in different latitudes (i.e.,  $ \rm JND_{lat} $), we also observed the relationship between 2D-JND and sample density, as shown in Fig.3. The relationship between 2D-JND and sample density approximately follows an exponential decay, as shown in Eq.(4).
\begin{equation}
	{\rm JND_{lat}} = ax^{b}+c.
\end{equation}

\begin{figure}[htpb]
	\vspace{-2mm}
	\centering
	\includegraphics[width=7.5cm,height=5cm]{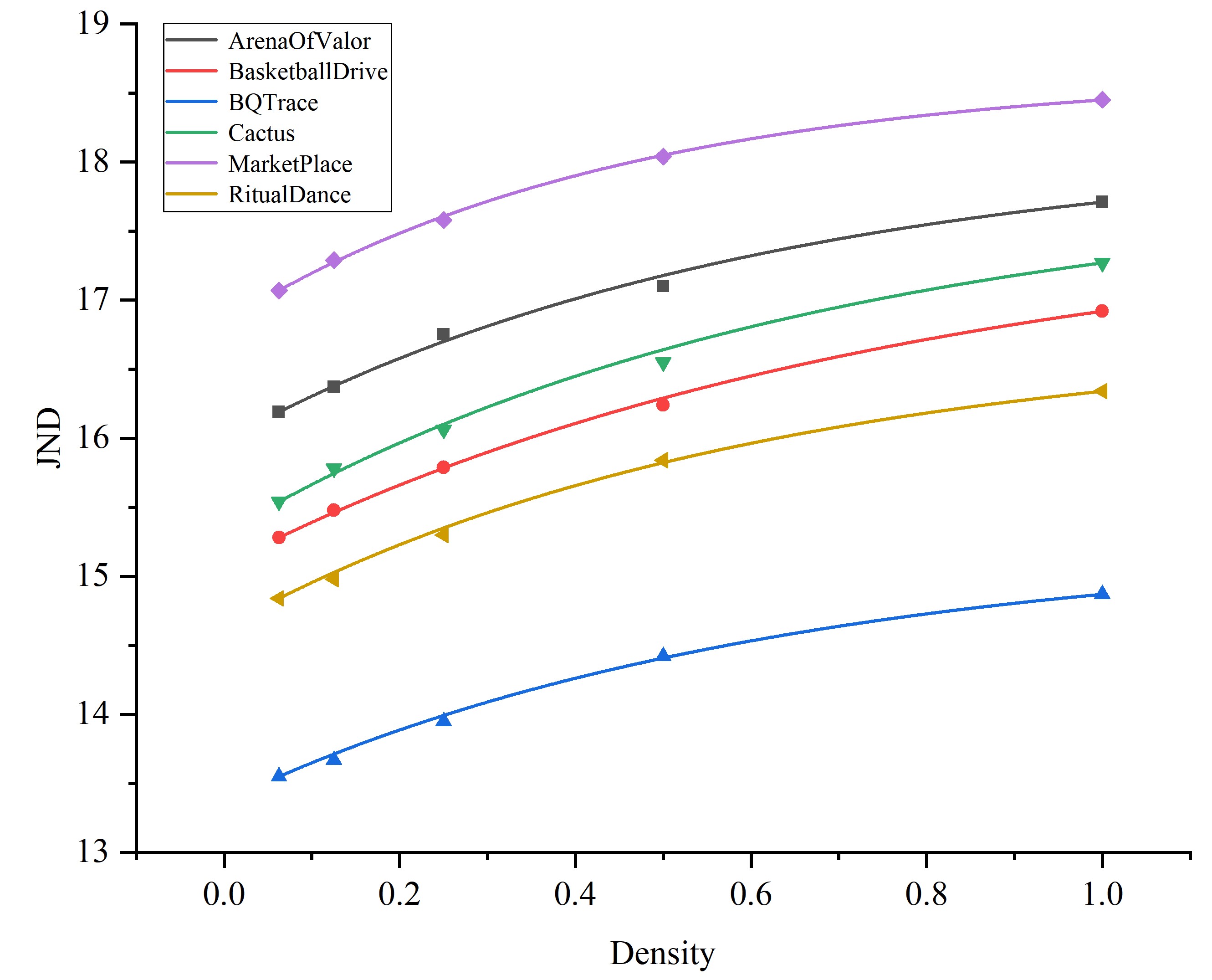}
	\caption{ Relationship Between 2D-JND and Sample Density.}
	\label{fig3}
\end{figure}

{\bfseries Latitude characteristic modeling:} Based on the relationship between $ \rm JND_{lat} $ and sample density (i.e.,Eq.(4)), we give the modeling process of the latitudinal characteristics. As shown in Fig.2(a), it can be seen that the perimeter of the sphere is equal to the width of the ERP panorama. For a given sample row in an ERP panorama, the number of samples $ M $ in each latitude-wise slice:
\begin{equation}
	M = round(2 \pi R),
\end{equation}
$ r $ denotes the radius of the sphere, $ R $ is the radius of the circle/latitude corresponding to the sample row in an ERP panorama and can be calculated as below:
\begin{equation}
	R = r \times \cos(\eta),
\end{equation}
$ \eta $  is the angle between the radius of the corresponding latitude and the radially connected sphere's center circle. This angle is given by Eq.(7):
\begin{equation}
	\eta = \cos((h - \frac{H}{2} + \frac{1}{2}) \times \frac{\pi}{H}),
\end{equation}
where $ h $ is the height of each pixel line, and $ H $ represents the height of an ERP image.Then, the amount of sample density change $ x $ can be defined by:
\begin{equation}
	x = \frac{W}{M},
\end{equation}

In summary, based on the $ \rm JND_{2D} $ model, the coefficients $ a $, $ b $ and $ c $ are obtained.
\begin{equation}
	\begin{array}{l}
		a = 0.13 \times \rm JND_{2D} - 0.11, \\[2mm]
		b = 0.034 \times \rm JND_{2D} + 0.27, \\[2mm]
		c = 0.86 \times \rm JND_{2D} + 0.21. \\
	\end{array}
\end{equation}

Finally, based on the Eq.(4), multiple $ \rm JND_{lat} $ curves constitute a set of curve clusters, as shown in Fig.4. The image of ERP projection has higher sampling density in the polar region, we observe the A point $ (x, \rm 2D\mbox{-}JND) $ is located in the curve cluster. Then, we make $ x=1 $ to obtain the $ \rm JND_{lat} $.
\begin{figure}[htbp]
	%\vspace{+4mm}
	\centering
	\includegraphics[width=7.5cm,height=5cm]{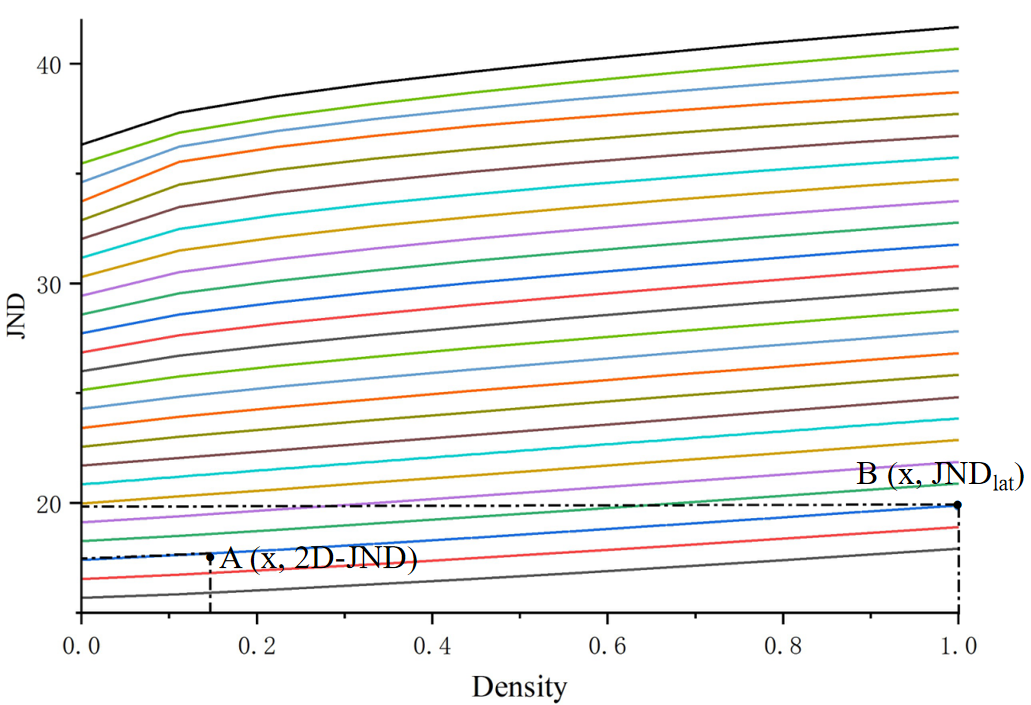}
	\caption{$ \rm JND_{lat} $ Curve Cluster.}
	\label{fig6}
\end{figure}

\subsection{Filed of View Modeling}
Due to the hardware limitations of the HMD, users can only view video content in the range of 90° to 120°, and more than half of it is invisible. Meanwhile, the HVS shows that users perceive image details with different visual sensitivities due to the uneven distribution of sensor cells on the retina. In general, as retinal eccentricity increases, visual acuity decreases and visibility threshold increases. Therefore, we first predict the interested regions of human eyes by attention models. Next, we investigate the effect of the different regions on spherical JND thresholds.

{\bfseries Filed of view characteristic:} Visual attention model aims to extract the region of visual interest in an image and generate a saliency map. When viewing a 360° video, only the content of projection images within the viewport is visible to the viewer. Therefore, we extract the viewport image and use the visual attention model to obtain the saliency map. Based on the introduction of human eye perspective characteristics, we assign 120° horizontal and vertical directions from the center of the sphere. Firstly, the longitude and latitude coordinates in ERP format are converted to 3D coordinates using the Cartesian Coordinates. Secondly, the three Euler angles, yaw(-180°to 180°), pitch(-90° to 90°) and roll(-90° to 90°), are utilized to obtain all possible viewport areas to be viewed. A partial display of the panorama as shown in Fig.5. 
\begin{figure}[htbp]
	\centering
	\includegraphics[width=7.5cm,height=4cm]{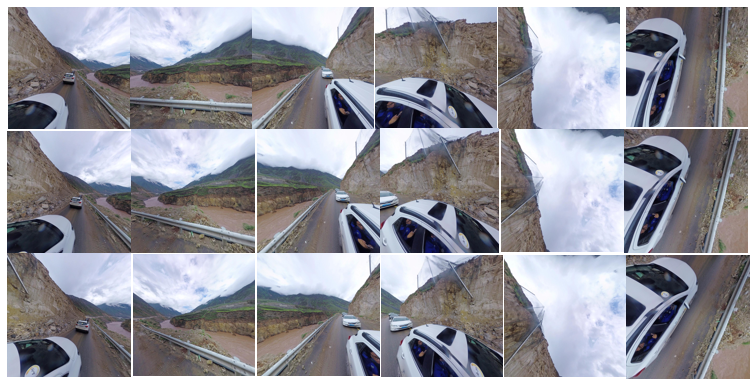}
	\caption{Viewport Display Regions.}
	\label{fig6}
\end{figure}

However, 360° videos are viewed differently from 2D videos. The traditional 2D image attention detection mechanism cannot be directly to 360° images.In recent years, saliency detection models for 360° videos have been widely studied. We select SalGAN360\cite{ref25} as our video saliency model based on comprehensive comparison and analysis of popular video saliency models. This model has better performance by applying fine tuning with new loss function, fusing the global and local saliency maps. The fusion of global and local image solves the problem that the global saliency map only roughly predicts the saliency of the equatorial region, while the local saliency map lacks globality. 

Foveation effect represents the phenomena that the fovea of retina has the highest visual acuity. It is a result of both the visual sensitivity and space-variant properties of the HVS. For given viewing area, distortions close to the area are more detectable, while distortions far from area are less detectable. Furthermore, part contents of 360° videos are not presented in the actual viewing, so it is necessary to consider the foveation effect. In addition, using a virtual reality HMD rather than a monitor to display 360° video scenes effectively solved the subject-to-screen distance problem and obtained a more accurate spherical JND threshold.

{\bfseries Filed of view modeling}: To explore the influence of field of view on the SJND threshold, we give the modeling process of the filed of view  characteristic. The HVS has the highest visual acuity at the region of area, the visual acuity decreases and the visibility threshold rises with increased retinal eccentricity. The contrast sensitivity function based on retinal eccentricity can be defined as:
\begin{equation}
	{\rm CT}(f, \tau) = {\frac{1}{64}} \times \exp(\chi f \frac {\tau+e_{1}} {e_{1}}),
\end{equation}
where $ f $ is spatial frequency, $ \tau $ is retinal eccentricity (degree). The experience values of $ \chi $ and $ e_{1} $ are 0.106 and 2.3, respectively.

Given the fixation point $ (i_{f}, j_{f}) $ and viewing distance $ v $, the retinal eccentricity of a $ (i, j)$-th point can be calculated as:
\begin{equation}
	\tau = min(\arctan( \frac {\sqrt {(i - i_{f})^2 + (j - j_{f})^2 }} {v}) .
\end{equation}

As can be seen form Fig.6, the gaze points of human eyes are not concentrated on a certain point of an image, but randomly distributed in a certain region, so the eccentricity is taken as the minimum distance from a certain point to a saliency region.Considering the directionality of human visual sensitivity, Eq.(14) can be modified as:
\begin{equation}
	{\rm CT}(f_{i, j}, \tau) = \frac {1/64} {r + (1 - r) \cos^2 \theta_{i,j}} \cdot \exp{\frac {\chi f_{i,j} (e_{1} + \tau)} {e_{1}} }, 
\end{equation}
\begin{equation}
	f_{i,j} = \frac {1} {2N \omega} \sqrt{i^2 + j^2},
\end{equation}
\begin{equation}
	\cos \theta_{i,j} = \frac {\left| i^2 - j^2 \right|} {i^2 + j^2},
\end{equation}
\begin{equation}
	\omega = \frac {\arctan(1/2D)} {W/2},
\end{equation}
where $ N $ refers to the dimension of a transform block, $ i $, $ j $ are the indexs of the $ (i, j) $-th DCT coefficient of a block. $ \omega $ is the corresponding frequency. $ D $ represents the ratio of viewing distance to $ W $.
\begin{figure}[htbp]
	\centering
	\includegraphics[width=7.5cm,height=4.5cm]{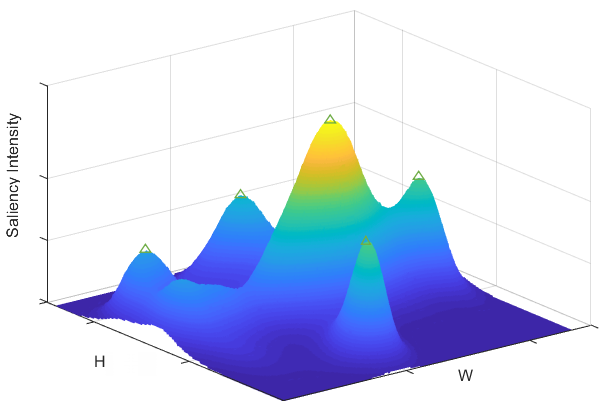}
	\caption{Human Eye Gaze Point Distribution Map.}
	\label{fig7}
\end{figure}

In addition, as shown in Fig.7(a), the eccentricity increases sharply as the distance from the peripheral point to the fixation point increases. However, the usual pattern of human visual foveation is that the fixation point moves slightly in a small area around the center point of interest, so the weighted value of foveation around the gaze point should be the same. The fixation point moves slightly in a small area around the center point of interest, so the weighted value of foveation around the gaze point should be the same. Therefore, to ensure the overall subjective perceptual quality, the acceleration of the visual conditioning factor should be reduced, as shown in Fig.7(b). Then, the weighting factor based on the foveation effect can be expressed as:
\begin{equation}
	\alpha_{fov}  = \left\{
	\begin{array}{cl}
		\ln \tau, & \mbox{$ \tau  \ge 2.7 $ and $ (i, j) \notin $ DC coefficient} \\[2mm]
		1, & others \\
	\end{array} , \right.
\end{equation}

\begin{figure}[htbp]
	\vspace{-4mm}
	\centering
	\subfloat[Traditional]{
		\includegraphics[width=4cm, height = 2.8cm]{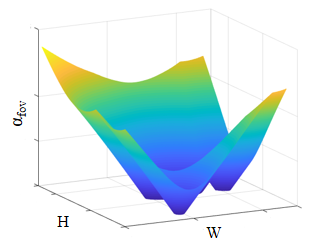}
		\label{a}
	}
	\subfloat[Proposed]{
		\includegraphics[width=4cm, height = 2.8cm]{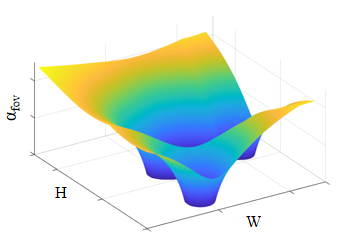}
		\label{c}
	}
	\caption{Foveation Weighting Factor.}
	\label{fig9}
\end{figure}

\subsection{The SJND Modeling}
Based on the above analysis, we extend our 2D-JND model to SJND by jointly exploiting latitude projection and field of view during 360° display. The proposed SJND model is illustrated in the Eq.(17). Fig.8 summarizes the proposed framework of SJND profile.
\begin{figure*}[t]
	\centering
	\includegraphics[width=17.5cm,height=4.6cm]{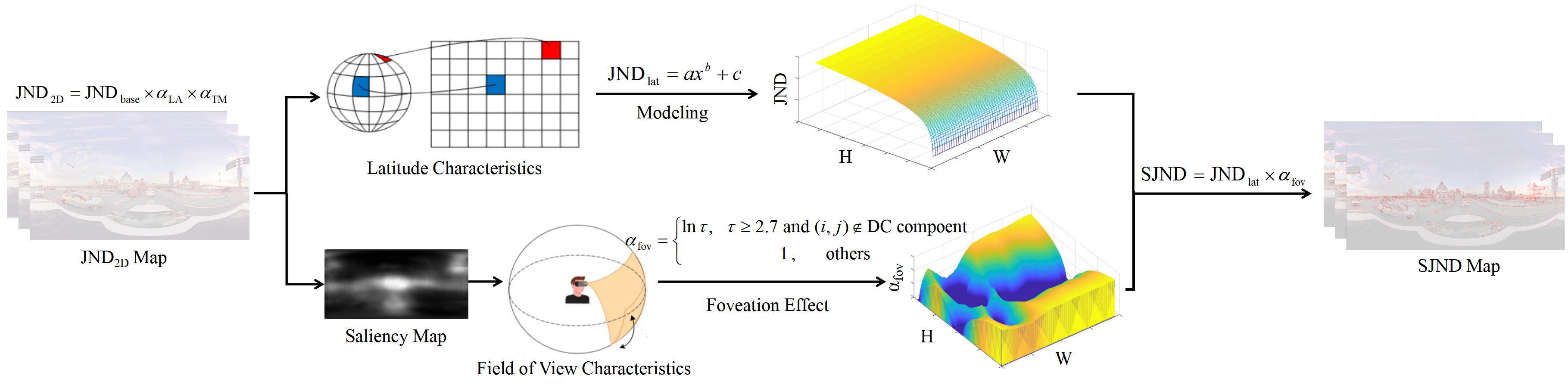}
	\caption{Framework of the Proposed Model for 360° videos.}
	\label{fig9}
\end{figure*}

\begin{equation}
	\begin{array}{l}
		{\rm JND_{2D}} = {\rm JND_{base}} \times \alpha_{\rm LA} \times \alpha_{\rm TM}, \\[2mm]
		{\rm JND_{lat} }= ax^{b}+c,  \\[2mm]
		{\rm SJND} = { \rm JND_{lat} }\times \alpha_{fov}. \\
	\end{array}
\end{equation}
where $ {\rm JND_{base}} $ and $ \alpha_{LA} $ are from \cite{ref12}, $ \alpha_{\rm TM} $ can be calculated from Eq.(3).

\section{Experiment Result}
In this section, 360° videos with different visual contents are chosen for testing, including 4K, 6K and 8K resolutions, as shown in Fig.9. To evaluate the performance of the JND models, noise is added to each DCT coefficient in an image according to the JND values.
\begin{equation}
	C^{'}(n, i, j) = C(n, i, j) + d \cdot {\rm SJND}(n, i, j),
\end{equation}
where $ C(n, i, j) $ is the $ (i, j)$-th DCT coefficient in the $n$-th block of an image. To avoid producing a fixed pattern of change, $ d $ is taken to be a random value of $ +1 $ or $ -1 $.
\begin{figure}[H]
	\centering
	\includegraphics[width=8.9cm,height=2.3cm]{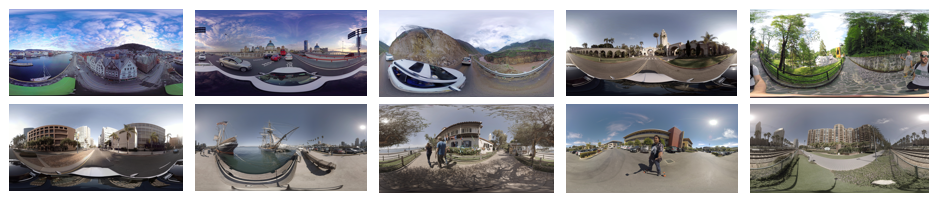}
	\caption{The Thumbnails of 10 Test Videos.}
	\label{fig10}
\end{figure}

\subsection{Performance Evaluation of proposed 2D-JND model}
To verify the superior performance of the proposed 2D-JND model, a comparison with the current state-of-the-art literature \cite{ref12}, \cite{ref15} and \cite{ref17} is presented. In general, a more accurate JND model will hide more noise with the same perception quality and the lower PSNR value. We set the SSIM value as 0.975 in the experiment to ensure the same perception quality.
\begin{table}[htpb]
	\centering
	\caption{PSNR Comparison of 2D-JND Model.}
	\renewcommand{\arraystretch}{1.2}
	\begin{tabular}{lcccc}
		\hline
		Sequence         & Ref.\cite{ref12} & Ref.\cite{ref15} & Ref.\cite{ref17} & 2D-JND         \\ \hline
		AerialCity       & 38.36       & 37.78      & 36.08        & \textbf{36.03} \\
		DrivingInCity    & 38.87       & 38.65      & 36.69        & \textbf{35.97} \\
		DrivingInCountry & 40.62       & 39.26      & 35.84        & \textbf{36.96} \\
		Balboa           & 39.81       & 37.28      & 35.92        & \textbf{35.28} \\
		BranCastle       & 40.89       & 38.51      & 37.43        & \textbf{36.78} \\
		Broadway         & 40.09       & 39.30      & 37.76        & \textbf{35.21} \\
		Harbor           & 41.11       & 41.06      & 39.60        & \textbf{38.70} \\
		KiteFlite        & 40.66       & 39.98      & 37.90        & \textbf{36.48} \\
		SkateboardInLot  & 41.58       & 41.23      & 40.59        & \textbf{40.10} \\
		Trolley          & 39.35       & 39.88      & 38.15        & \textbf{37.70} \\
		Average          & 40.13       & 39.29      & 37.60        & \textbf{36.92}
		\\ \hline
	\end{tabular}
\end{table}

As can be seen from Table 3, our proposed 2D-JND model has the lowest PSNR value with an average of 36.37dB. Compared with the existing JND models, our average PSNR decreased by 3.21dB, 2.37dB and 0.68dB.This further illustrates the superiority of the proposed 2D-JND model.

To reflect the distribution of JND thresholds more intuitively, Fig.10 gives the JND threshold distributions of some test sequences. It is obvious from the figures that Wei's model introduces a lot of noise to flat regions, such as in the sky and face regions in the figure. In addition, the detection of edge regions is not accurate, such as the edge of the fence in the BranCastle video frame and the obvious edges of houses such as the roofs in the KiteFlite and AerialCity video frames are not accurately distinguished. Wu's model and Wang's model can detect the edges better than Wei's model, which assign a huge amount of noise in the texture complex region, but still overestimates the JND threshold for flat regions such as the sky and lake in the image. In comparison, our 2D-JND model can effectively balances the JND distribution in flat, edge and texture regions, avoiding the introduction of noise in flat regions and introducing more noise into texture complex regions. Thus, the experiment results verify the performance of the proposed information entropy-based block classification method, further illustrating that our 2D-JND model is more accurate.
\begin{figure*}[htpb]
	\centering
	\vspace{3mm}
	\begin{minipage}{0.24\linewidth}
		%\vspace{3pt}
		\centerline{\includegraphics[width=4.1cm, height = 2.2cm]{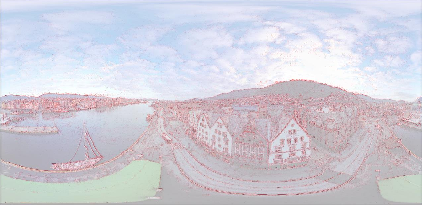}}
		\centerline{(a)AerialCity($ \rm JND_{2D}$)}
	\end{minipage}
	\begin{minipage}{0.24\linewidth}
		\centerline{\includegraphics[width=4.1cm, height = 2.2cm]{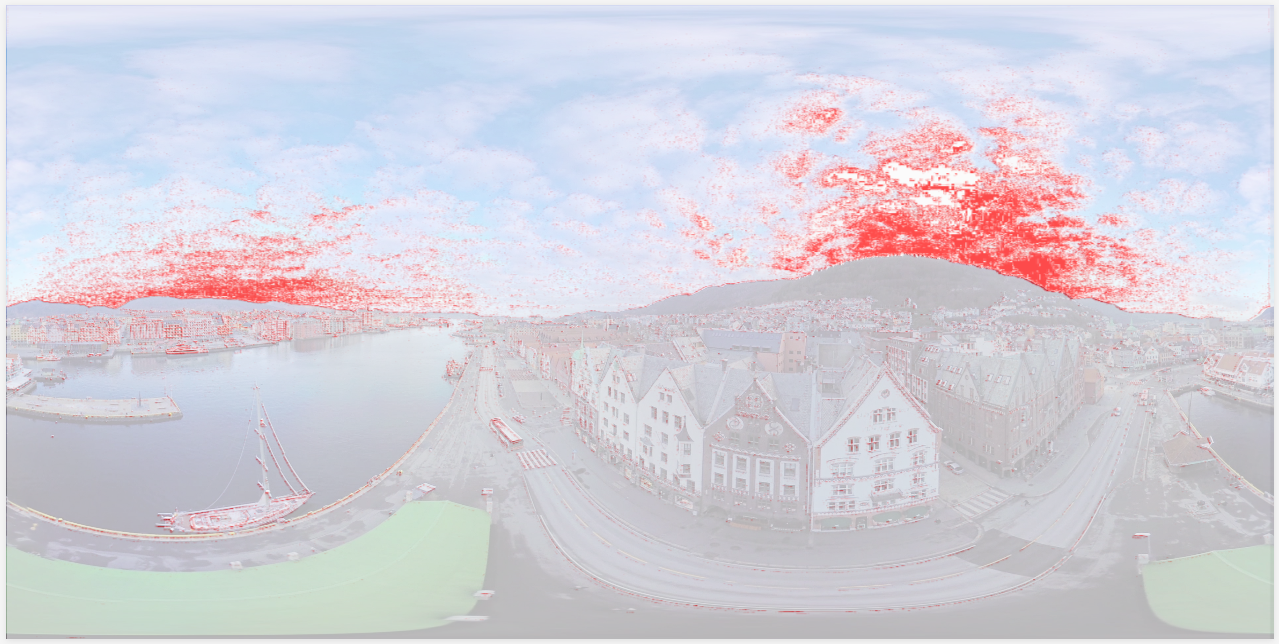}}
		\centerline{(b)AerialCity(Wei)}
	\end{minipage}
	\begin{minipage}{0.24\linewidth}
		\centerline{\includegraphics[width=4.1cm, height = 2.2cm]{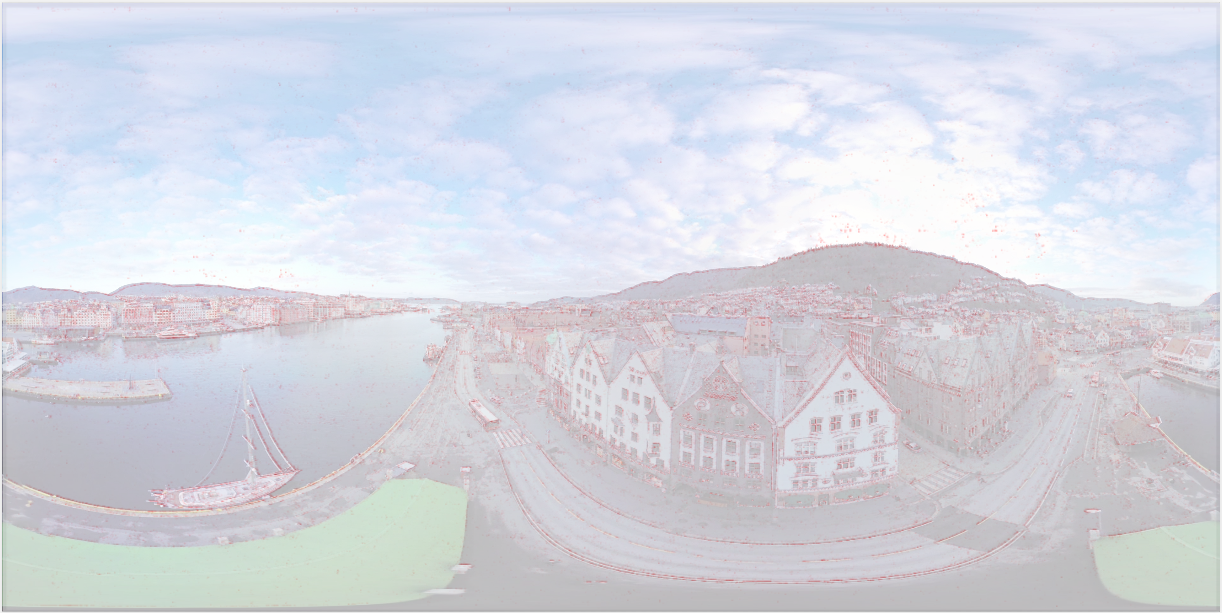}}
		\centerline{(c)AerialCity(Wu)}
	\end{minipage}
	\begin{minipage}{0.24\linewidth}
		\centerline{\includegraphics[width=4.1cm, height = 2.2cm]{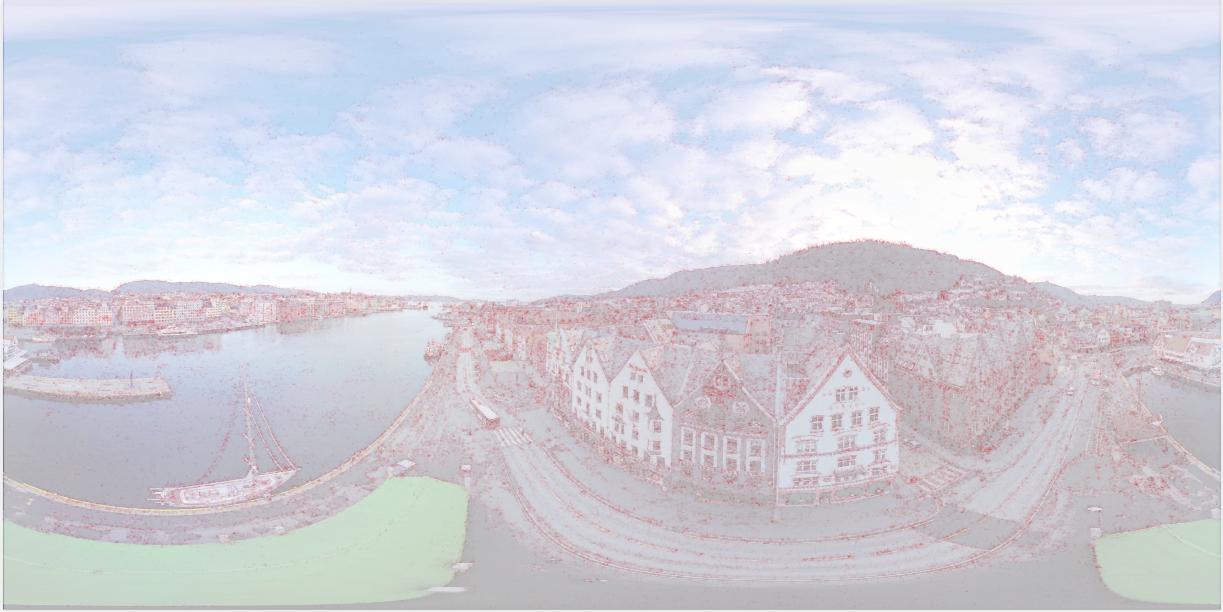}}
		\centerline{(d)AerialCity(Wang)}
	\end{minipage}	
	\begin{minipage}{0.24\linewidth}
		\centerline{\includegraphics[width=4.1cm, height = 2.2cm]{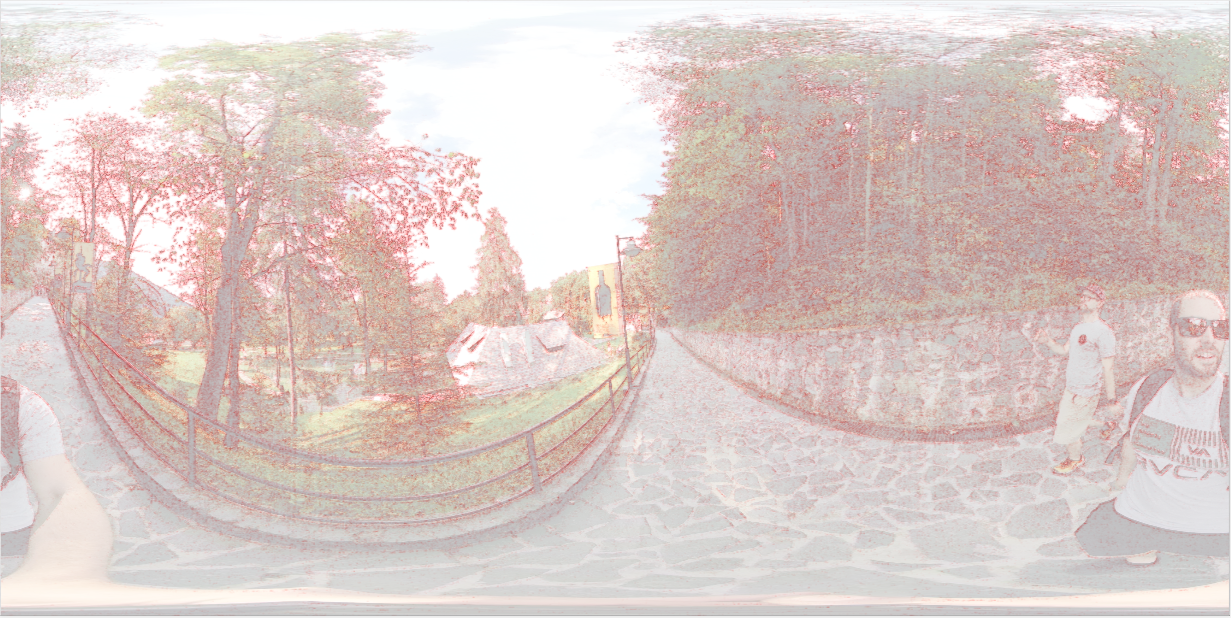}}
		\centerline{(a)BranCastle($ \rm JND_{2D}$)}
	\end{minipage}
	\begin{minipage}{0.24\linewidth}
		\centerline{\includegraphics[width=4.1cm, height = 2.2cm]{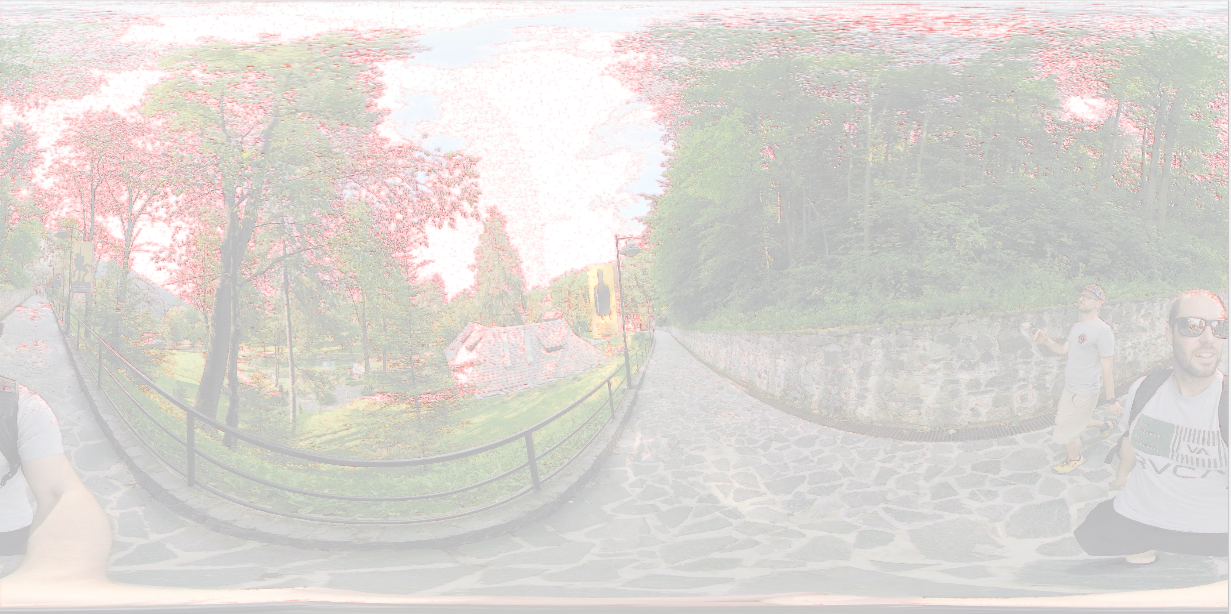}}
		\centerline{(b)BranCastle(Wei)}
	\end{minipage}
	\begin{minipage}{0.24\linewidth}
		\centerline{\includegraphics[width=4.1cm, height = 2.2cm]{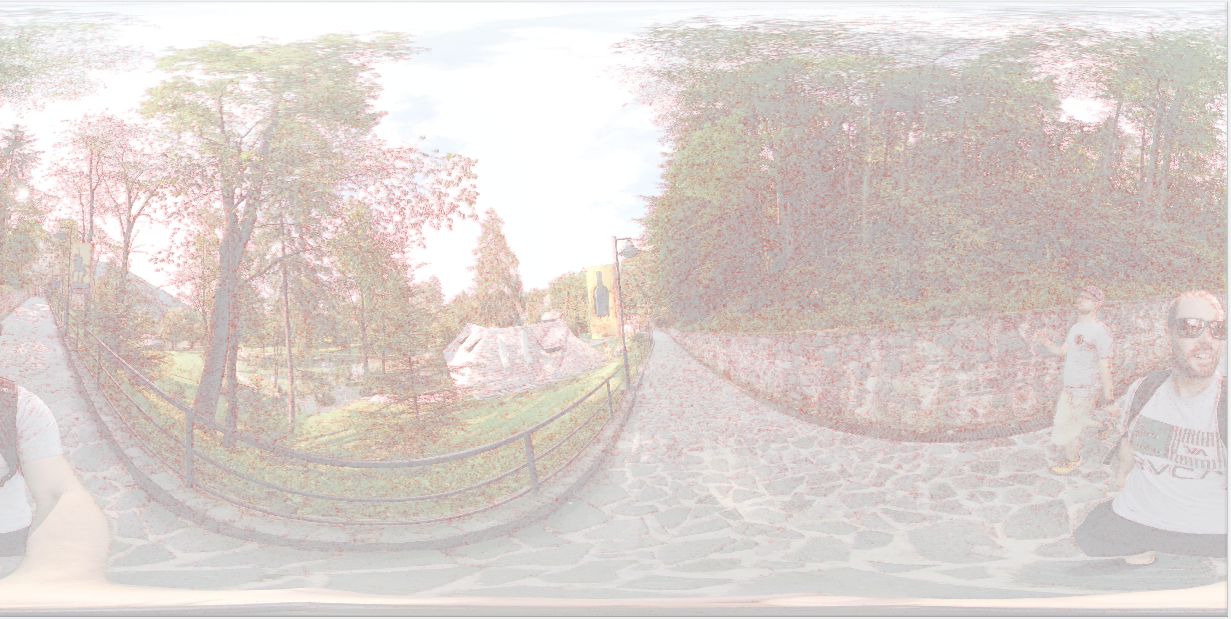}}
		\centerline{(c)BranCastle(Wu)}
	\end{minipage}
	\begin{minipage}{0.24\linewidth}
		\centerline{\includegraphics[width=4.1cm, height = 2.2cm]{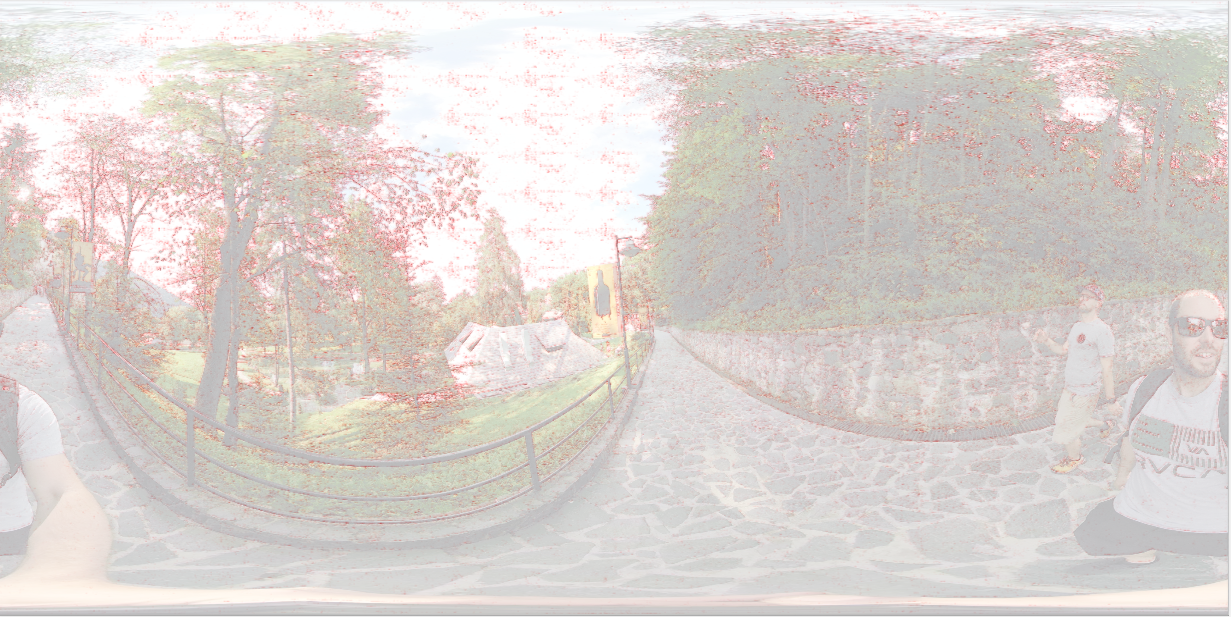}}
		\centerline{(d)BranCastle(Wang)}
	\end{minipage}
	\begin{minipage}{0.24\linewidth}
		\centerline{\includegraphics[width=4.1cm, height = 2.2cm]{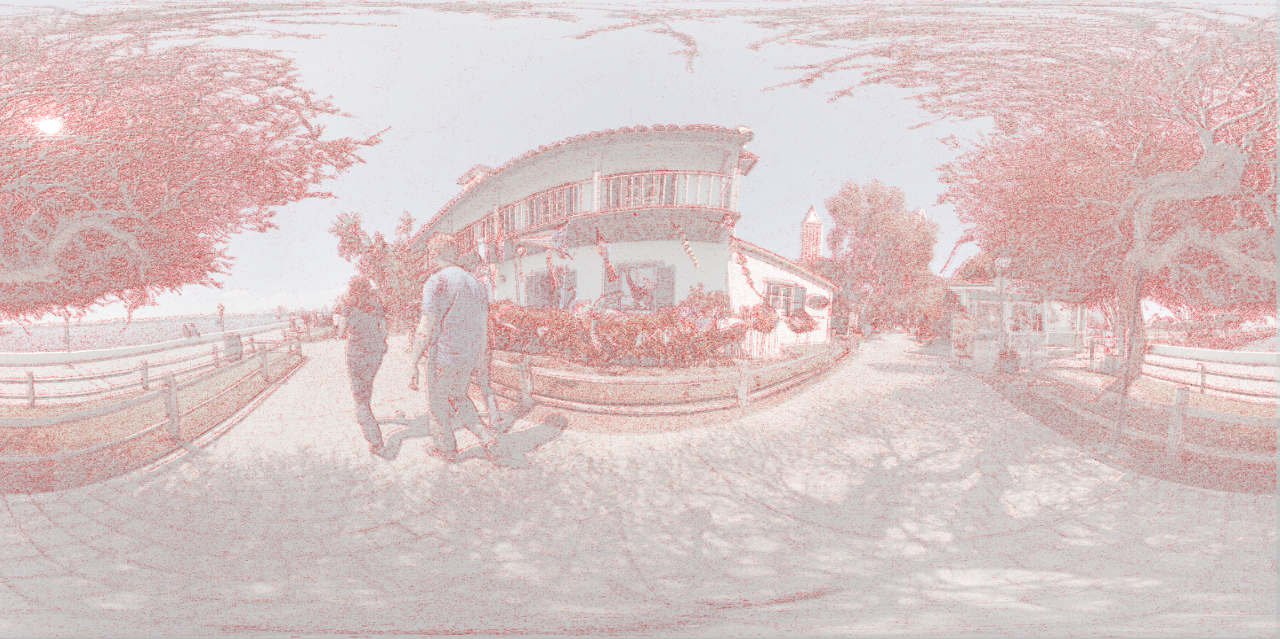}}
		\centerline{(a)KiteFlite($ \rm JND_{2D}$)}
	\end{minipage}
	\begin{minipage}{0.24\linewidth}
		\centerline{\includegraphics[width=4.1cm, height = 2.2cm]{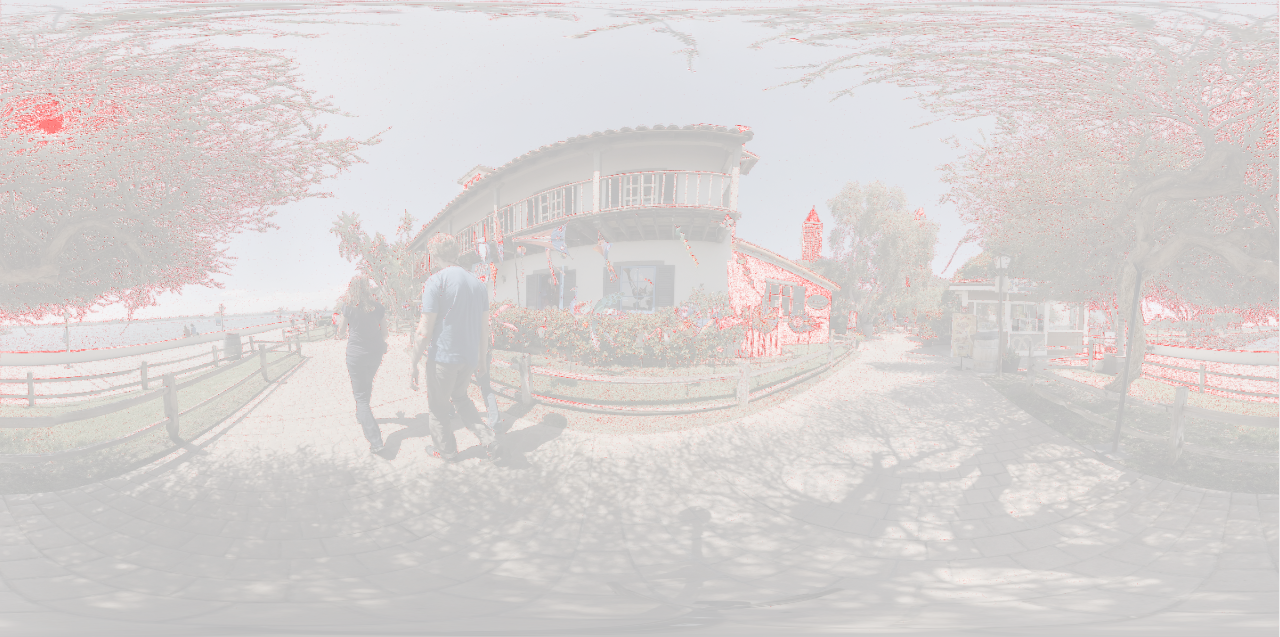}}
		\centerline{(b)KiteFlite(Wei)}
	\end{minipage}
	\begin{minipage}{0.24\linewidth}
		\centerline{\includegraphics[width=4.1cm, height = 2.2cm]{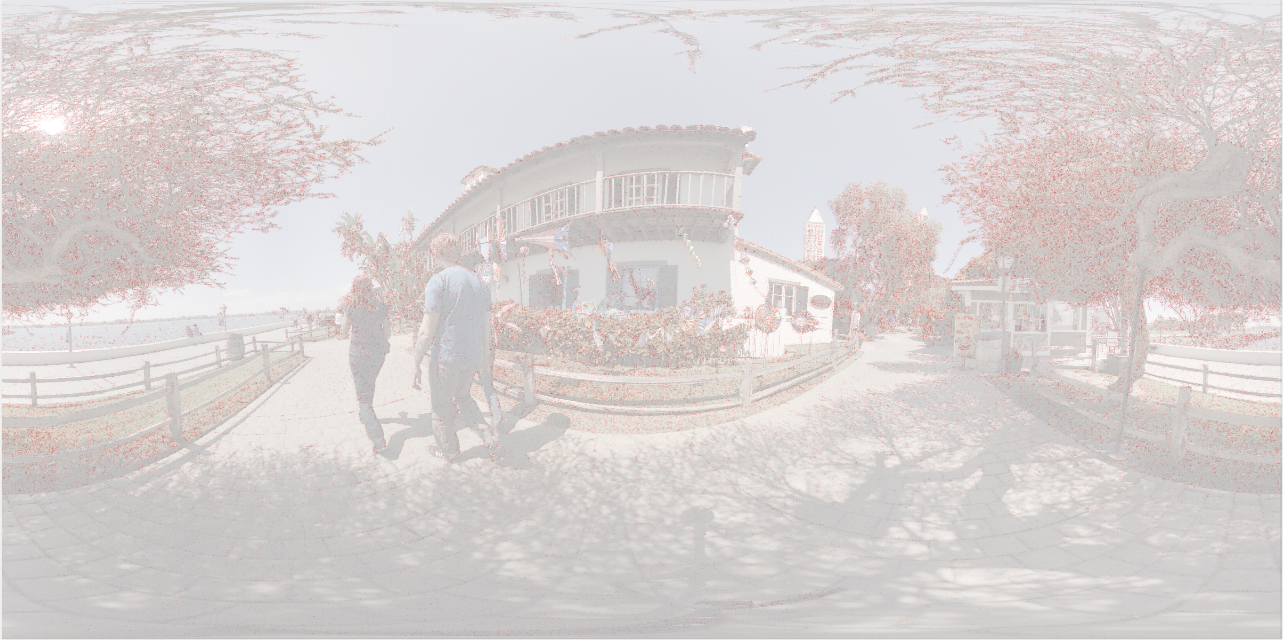}}
		\centerline{(c)KiteFlite(Wu)}
	\end{minipage}
	\begin{minipage}{0.24\linewidth}
		\centerline{\includegraphics[width=4.1cm, height = 2.2cm]{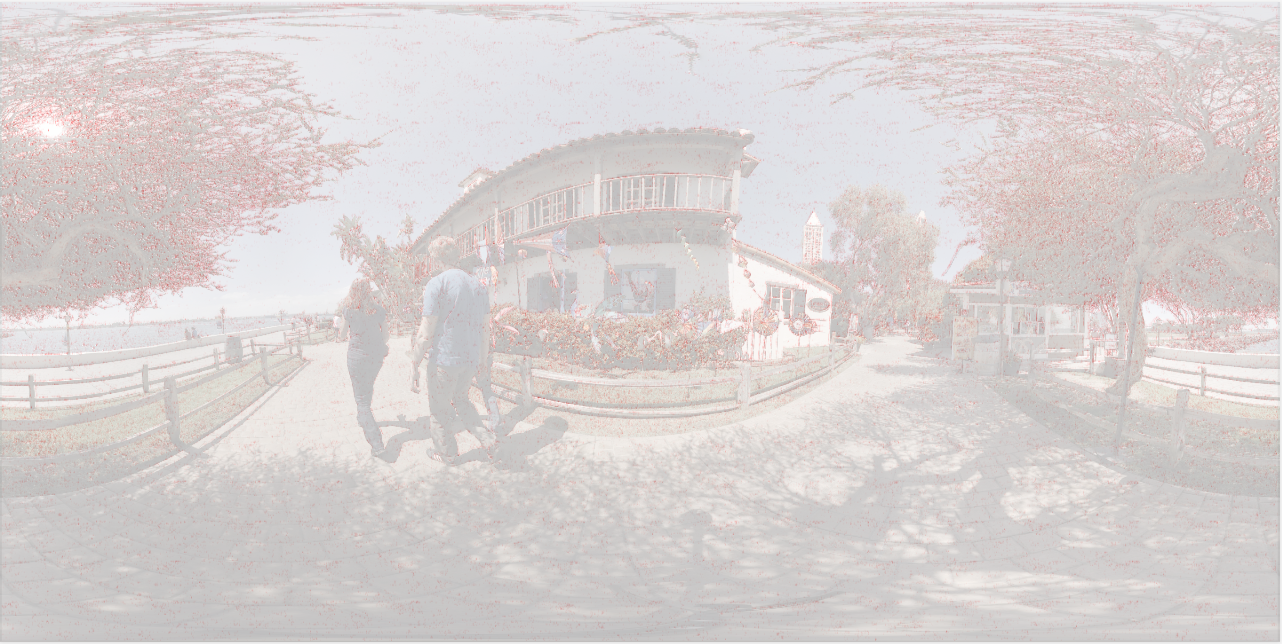}}
		\centerline{(d)KiteFlite(Wang)}
	\end{minipage}
	
	\caption{$ \rm JND_{2D} $ Threshold Maps:(a) JND threshold distributions of our proposed $ \rm JND_{2D} $; (b) JND threshold distributions of \cite{ref12}; (c) JND threshold distributions of \cite{ref15}; (d) JND threshold distributions of \cite{ref17}. }
	\label{fig11}
\end{figure*}

\subsection{Subjective Quality Experiments of SJND Model}
We set a subjective quality experiment for determining whether users can perceive the quality difference between original video and the video sequence injected with JND noise. 

\begin{itemize}
	 \item {\bfseries Experimental environment settings:} The head-mounted display is HTC VIVE, which has a binocular 110° 2K HD field-of-view range and refresh rate is 90 Hz. The test is executed in a laboratory environment with a normal indoor lighting level. During the process, lighting remains constant and the environment keeps silent. The above conditions are set and calibrated according to recommendations of the ITU-R BT.500-11\cite{ref26}. 
	
	\item {\bfseries Test data:} We employ 20 standard 360° video pairs as shown in Fig. 10 as materials, each video pair contains an original video and a video with JND noise, correspondingly.  
	
	\item {\bfseries Experiment:} In the subjective testing, we invite 21 subjects, including 12 males and 9 females. Subjects are shown 10 video groups randomly. Each subject is invited to watch two video sequences in turn by wearing HTC VIVE device and judges whether there are any differences in quality between video sequences. Specifically, source video and video injected with JND noise of a video pair are presented in random order. There is a 3-s gray screen between two video pairs to prompt subjects to make a judgment. Finally, a consistency check is conducted for each subject by randomly selecting 5 repeated video pairs out of 20 pairs. A subject is rejected if the number of different choices given to repeated pairs is more than three. As a result, no subjects is rejected.
\end{itemize}

\begin{figure}[H]
	\centering
	\includegraphics[width=9cm,height=7.5cm]{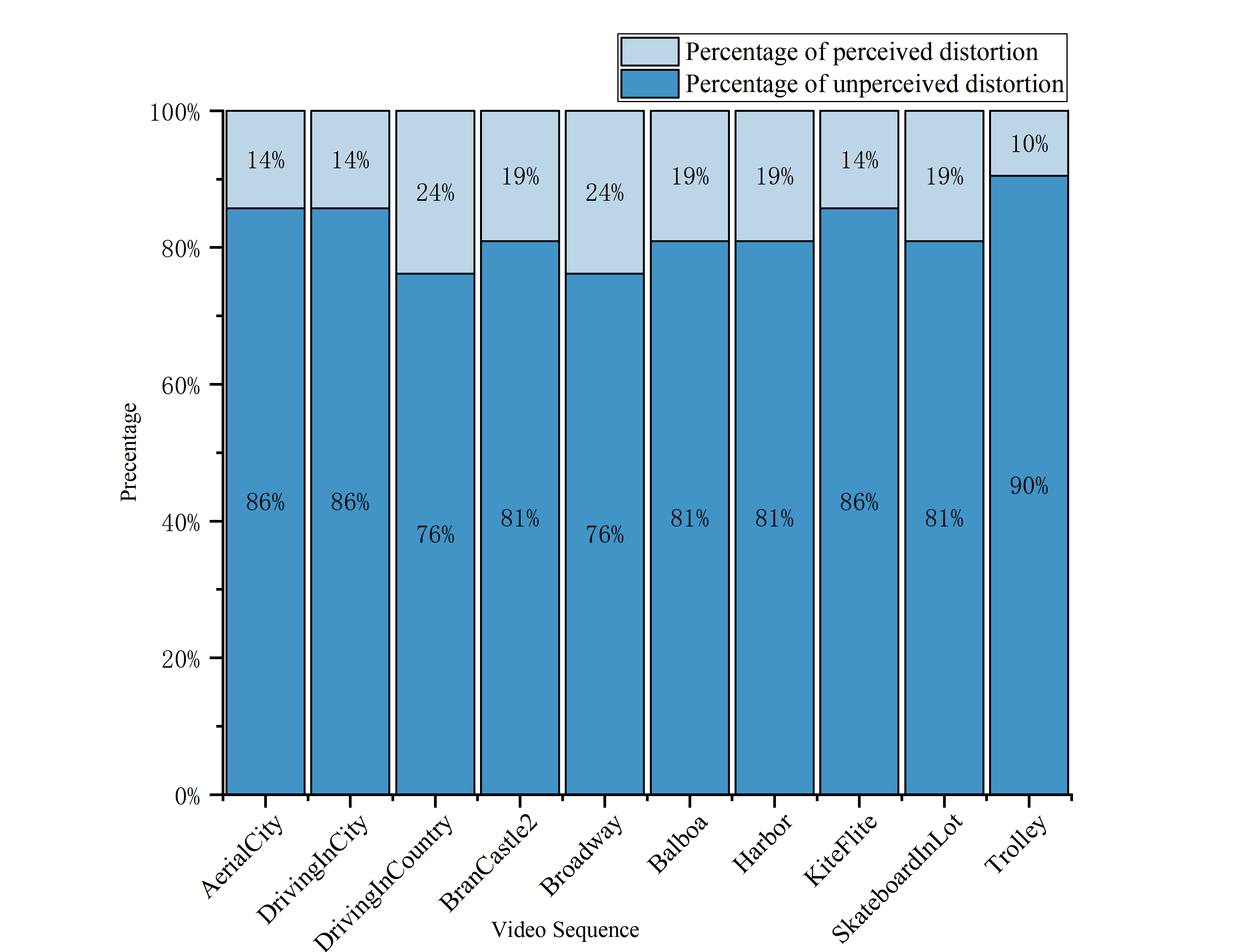}
	\caption{The Bar Graph of Subjective Experimental Results.}
	\label{fig11}
\end{figure}

The experimental results are exhibited in Fig.11, where the $ x $-axis is the name of a video pair, the $ y $-axis represents the proportion of subjects who do not perceive differences in quality of video pairs. It can be observed that all the test sequences quality satisfy more than 75\% of subjects cannot distinguish quality differences, which further illustrates the proposed SJND model can effectively hide the distortion without compromising perceived quality.

\subsection{Objective Quality Experiments}
To further verify the performance of the proposed SJND model, we measure the performance in terms of PSNR for test videos. At the same perceptual quality, the higher injected-noise, the PSNR shows a lower values. It means the SJND profile can effectively hide the distortion and has a good performance compare with other models. Our proposed SJND model is compared with the state-of-art and unique panoramic video JND model\cite{ref27} for validation.The comparison results of the objective quality assessment experiments are shown in Table 4.
\begin{table}[htpb]
	\centering
	\caption{PSNR Comparison of SJND Model.}
	\renewcommand{\arraystretch}{1.2}
	
	\begin{tabular}{lcc}
		\hline
		&Ref.\cite{ref27}  &SJND   \\ \hline
		AerialCity           & 35.50          & \textbf{33.75}            \\
		DrivingInCity        & 37.54          & \textbf{35.39}            \\
		DrivingInCountry     & 37.17       & \textbf{35.06}             \\
		BranCastle2         & 35.11          & \textbf{33.66}           \\
		Broadway            & 37.29          & \textbf{35.82}            \\
		Balboa              & 37.18      & \textbf{32.95}           \\
		Harbor               & 38.32          & \textbf{36.13}          \\
		KiteFlite          & 37.73      & \textbf{34.32}           \\
		SkateboardInLot      & 38.25         & \textbf{35.64}             \\
		Trolley             & 37.74        & \textbf{34.13}            \\
		Average              & 37.18          & \textbf{34.69}        \\
		\hline
	\end{tabular}
\end{table}

Experimental results suggests that the average PSNR of our SJND model is 34.69dB, which is 2.49dB lower than Ref.\cite{ref27} on average. The experimental results further illustrate that the proposed SJND model effectively exploits HVS characteristics and  360° video features, which allowing more distortion to be tolerated while ensuring the subjective quality and more accurate estimation of visual redundancy in immersive scenes. The surperiority of the SJND model can be summarized as follows.

On the one hand, the traditional 2D JND models are for 2D image contents, which are lack of the research on the characteristics of 360° videos. Although the Sami's model is designed for 360° image, it fails to develop the field of view of 360° video scenes. According to Fig.12, we investigate the impact of sampling density on JND modeling and provide the equator region greater JND values based on the features of ERP format. On the other hand, As shown in Fig.13, we introduced larger weighting coefficients for regions far from saliency. Also, we utilize the attention mechanism of HVS and the foveation effect to investigate the effect of visual field on JND modeling.Therefore, the proposed SJND model shows remarkable performance than another models.
\begin{figure}[htpb]
	%\vspace{-3mm}
	\centering
	\includegraphics[width=8.4cm, height = 2.7cm]{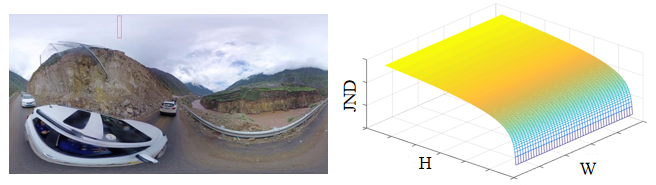}
	\caption{Distribution of JND Values in Different Latitude Regions for 360° videos.}
	\label{fig12}
\end{figure}

\begin{figure}[htpb]
	\vspace{-8mm}
	\centering
	\includegraphics[width=8.4cm, height = 2.7cm]{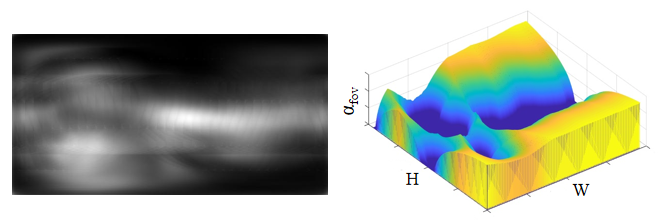}
	\caption{Weight Distribution of Different Attention Areas of 360° videos.}
	\label{fig13}
\end{figure}

In order to further verify that it is necessary to consider the latitude and field of view characteristics of 360° videos, we show the enlarged images of partial details for different latitude regions and viewport regions of 360° videos in the experiment, as shown in Fig.14. In the plain regions such as the sky and lake, Ref.\cite{ref27} overestimates the visual threshold and introduces a large amount of noise, while the proposed SJND model assigns more noise to regions such as lawns and woods, which is not sensitive for the HVS. Besides, the more noise is introduced with increasing latitude. In the regions with high human attention such as license plates, street lights, and car, Ref.\cite{ref27} is lack of considering the visual attention mechanism and the field-of-view range, resulting in poor the subjective perception.
\begin{figure*}[t]
	%\vspace{-13.58mm}
	\centering
	\subfloat[Original.]{
		\includegraphics[width=18.0cm, height = 3.4cm]{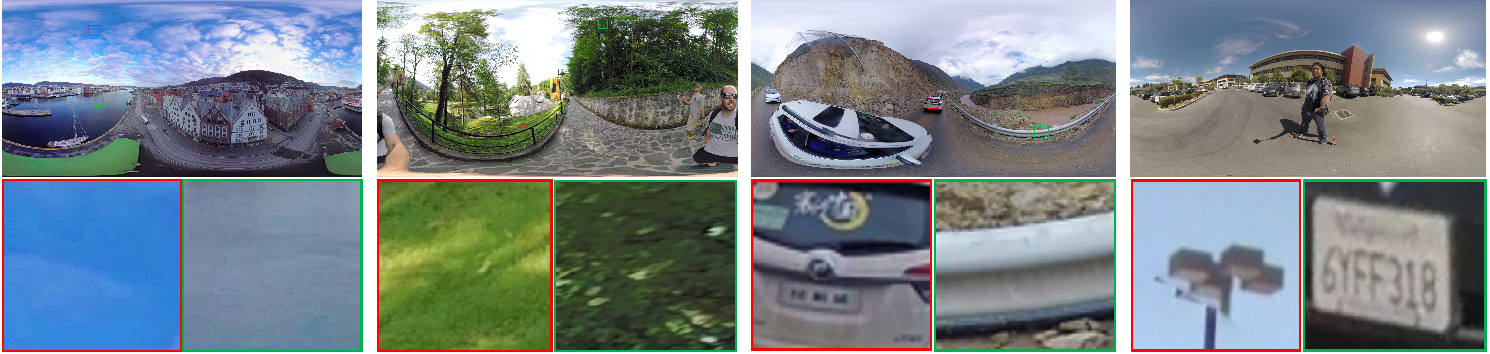}
	}
	\quad
	\subfloat[SJND Model.]{
		\includegraphics[width=18.0cm, height = 3.4cm]{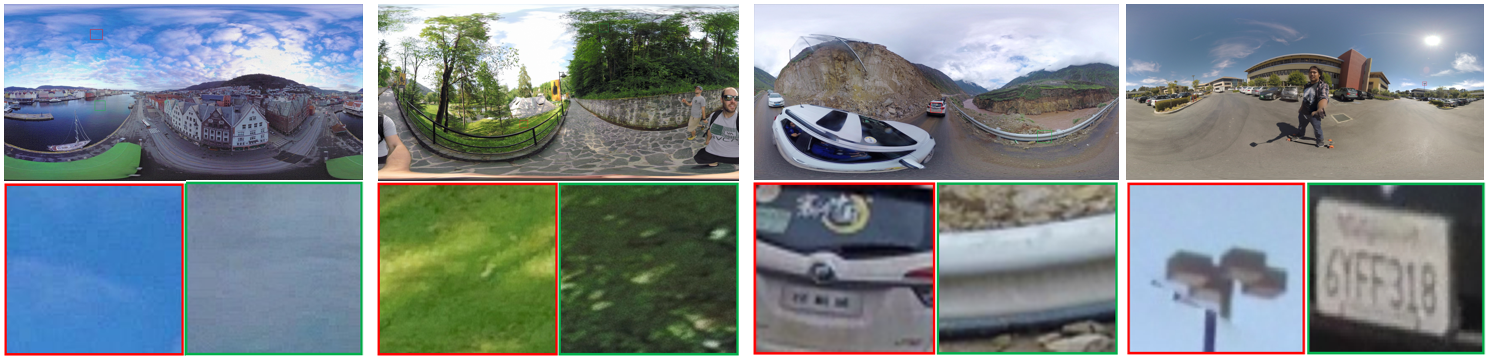}
	}
	\quad
	\subfloat[Ref.\cite{ref27}]{
		\includegraphics[width=18.0cm, height = 3.4cm]{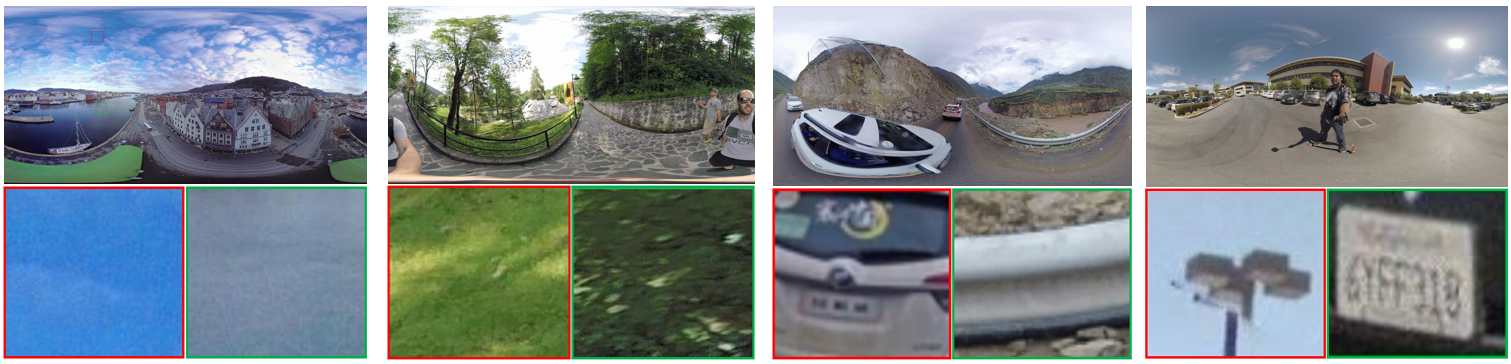}
	}
	\caption{Enlarged View of Local Details:(a) detail enlarged of original images; (b) detail enlarged of our SJND model; (c) detail enlarged of \cite{ref27}.}
	\label{fig14}
\end{figure*}

\subsection{Ablation Study}
Since the field of view characteristics and the visual attention of  360° videos have a great impact on the subjective perception  and will further improve the coding efficiency, we design an ablation experiment to verify the influence of human visual attention on the coding algorithm, as shown in Table 6. The SJND model is compared with the JND model without considering the weighting factor of the field-of-view  of 360° videos. The test conditions remained the same as before.
\begin{table}[htpb]
	\centering
	\renewcommand{\arraystretch}{1.2}
	\caption{Coding Performance Comparison.}
	\begin{tabular}{llcc}
		\hline
		Resolution          & Sequence         & JND    & SJND-VVC        \\ \hline
		\multirow{3}{*}{4K} & AerialCity       & -2.29\% & \textbf{-2.98\%} \\
		& DrivingInCity    & -2.13\% & \textbf{-3.10\%} \\
		& DrivingInCountry & -2.48\% & \textbf{-4.13\%} \\
		\multirow{3}{*}{6K} & Balboa           & -2.09\% & \textbf{-5.21\%} \\
		& BranCastle       & -3.08\% & \textbf{-4.49\%} \\
		& Broadway         & -2.50\%  & \textbf{-4.12\%} \\
		\multirow{4}{*}{8K} & Harbor           & -1.97\% & \textbf{-3.51\%} \\
		& KiteFlite        & -3.05\% & \textbf{-3.70\%} \\
		& SkateboardInLot  & -2.17\% & \textbf{-5.06\%} \\
		& Trolley          & -2.56\% & \textbf{-3.50\%} \\
		Average             &                  & -2.43\% & \textbf{-3.98\%}  \\ \hline
	\end{tabular}
\end{table}

The experimental results show that when the visual attention and visual centrality effects of 360° videos are not considered, the BD-Rate still improves compared with the original encoder, but decreases compared with the proposed SJND model, with an average decrease of 1.55\%. The reason is that the SJND model incorporating visual attention mechanism and visual foveation effect can obtain the minimum visual threshold of human eyes to ensure the video quality. On the other hand, the visual attention mechanism and visual foveation effect can be used to express the importance of visual contents and the distribution of human eyes' attention level to reasonably allocate bit resources. Therefore, combining the two can give full play to their respective advantages to reduce the coding bit rate and improve the coding efficiency while guaranteeing the quality.

\section{The Application of The SJND Model in Video Coding}
To further examine the effectiveness of our SJND model, we embed it in VVC compression. In addition, we also analyze the performance of its experimental results.

\subsection{Video coding based on the SJND model}
The current mainstream video coding standards mainly use a hybrid coding framework. In this framework, an original video is predicted and the residuals are compressed by transform, quantization and entropy coding. In the quantization process, the Quantization Parameter (QP) plays a key role. In fact, the QP reflects the compression of spatial details. A smaller QP means that most of the details are retained. Otherwise, a larger QP means that some details are lost, the bit rate is reduced and the quality of the image is degraded. The traditional QP used for residual DCT coefficient quantization or inverse quantization can be expressed as:
\begin{equation}
	\rm QP = QP_{0} + \Delta Q
\end{equation}
where $ {\rm QP_{0}} $ is an original quantization parameter of current coding units and used for uniform quantization in a coding unit. The QP also fails to fully explore the perceptual properties under optimal mode coding. According to the SJND model, different CUs have different visibility thresholds, a higher SJND threshold indicates that more distortion can be tolerated. The QP of a CU block can be set to a larger value to reduce the coded bits while ensuring subjective perception. Conn with a lower SJND threshold, a smaller QP is utilized to ensure video quality. Therefore, the improved QP b
Similar to \cite{ref28, ref29}, the perceptual QP for each CU is defined as follows:
\begin{equation}
	{\rm QP_{CU}} = {\rm QP_{0}} \times \sqrt{m + \frac {n}{1 + p \times \exp(-q \times \frac{\bar{J}_{\rm CU} - \bar{J}_{\rm frame}}{\bar{J}_{\rm frame}})}},
\end{equation}
where $ \rm QP_{CU} $ is the perceptual QP of a CU block. $ \bar{J}_{\rm CU} $ and the $ \bar{J}_{\rm frame} $ are the average values of the current CU and frame, respectively. $ m , n, p, q $ are constants.

\subsection{Performance Analysis}
In order to evaluate the performance of our proposed SJND model for encoding, comprehensive experiments are implemented on the top of 360Lib-11.0 software based on the VVC (VTM10.0). A total of ten test video sequences included three different resolutions of 4K, 6K and 8K. The experiments are conducted under the common test conditions (CTC)\cite{ref30}.
\begin{table}[htpb]
	\centering
	\caption{Performance Comparison of BD-rate.}
	\renewcommand{\arraystretch}{1.2}
	\resizebox{82mm}{21mm}{
		\begin{tabular}{llcccc}
			\hline
			Resolution  & Sequence  & SJND-VVC    &Ref.\cite{ref27}    &Ref.\cite{ref17}  &Ref.\cite{ref27}     \\ \hline
			\multirow{3}{*}{4K}  & AerialCity     & \textbf{-2.98\%}     & -1.33\%      & -2.09\%     & -2.67\%     \\
			& DrivingInCity     & \textbf{-3.10\%} & -1.04\%   	& -1.82\%			& -2.03\%	\\
			& DrivingInCountry  & \textbf{-4.13\%}     & -0.91\%      & -2.71\%      & -1.18\%        \\ 
			\multirow{3}{*}{6K}  & Balboa    & \textbf{-5.21\%}     & -1.05\%    & -2.05\%   & -2.78\%            \\
			& BranCastle      & \textbf{-4.49\%}     & -0.89\%    & -1.38\%      	& -1.74\% 	\\
			& Broadway        & \textbf{-4.12\%}     & -1.20\%   & -1.98\%			& -2.17\%	\\
			\multirow{4}{*}{8K}  & Harbor  & \textbf{-3.51\%}     & -1.54\%   	& -2.37\%	& -2.66\%		    \\
			& KiteFlite     & \textbf{-3.70\%}     & -1.92\%   	& -2.14\%		& -3.22\%	    \\
			& SkateboardInLot   & \textbf{-5.06\%}     & -1.67\%  	& -2.16\%			 & -1.94\%	\\
			& Trolley         & \textbf{-3.50\%}     & -1.33\%  	& -2.03\%		 & -2.26\%	\\
			Average & \multicolumn{1}{c}{\textbf{}} & \textbf{-3.98\%}     & -1.29\%   	&-2.07\%		& -2.27\%	\\ \hline
		\end{tabular}
	}
\end{table}

Table 5 indicates that the proposed adaptive QP can consistently outperform the 360Lib9.0-VTM10.0 anchor and another scheme. The BD-rate is reduced by 3.98\% on average. This is because the video content is characterized by the existence of large texture complexes, the lower-level regions account for the majority of the entire video content, while the regions with higher texture complexity are extremely complex. In this way, the CTUs with high texture complexity are adaptively increased the QP while ensuring the video quality, and the bite rate is greatly reduced.  
\begin{figure*}[htpb]
	\centering
	\vspace{-3mm}
	\subfloat[DrivingInCity(VVC)]{
		\includegraphics[width=8.5cm, height = 3.3cm]{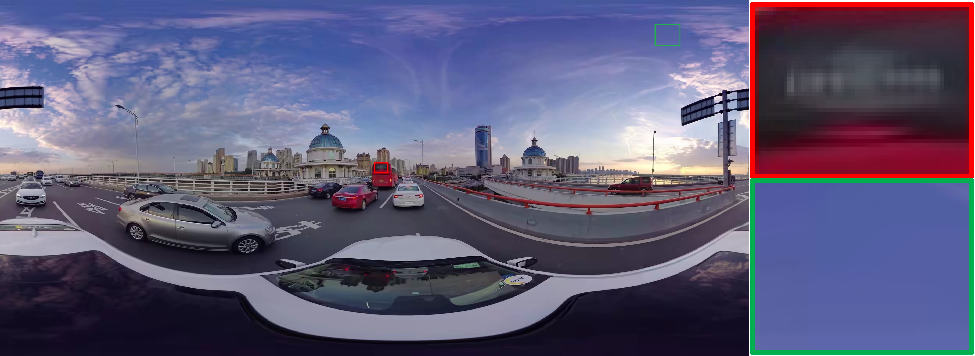}
	}
	\quad
	\subfloat[DrivingInCity(SJND-VVC)]{
		\includegraphics[width=8.5cm, height = 3.3cm]{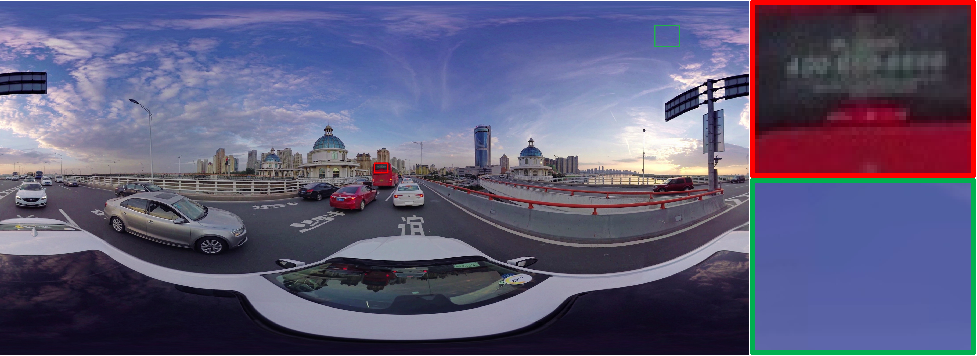}
	}
	\quad
	\subfloat[DrivingInCountry(VVC)]{
		\includegraphics[width=8.5cm, height = 3.3cm]{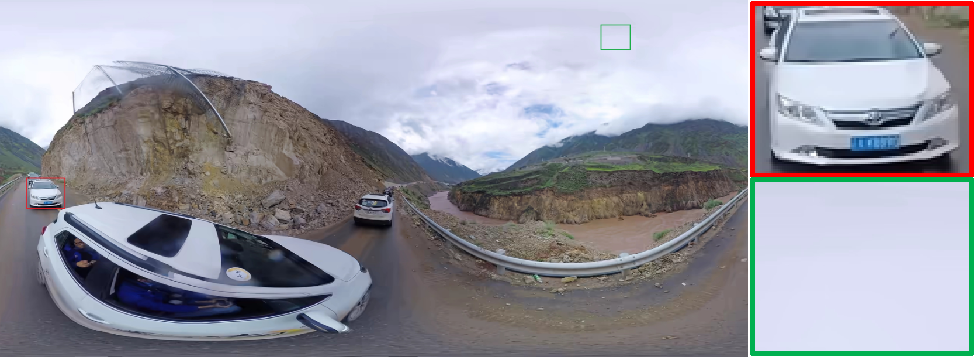}
	}
	\quad
	\subfloat[DrivingInCountry(SJND-VVC)]{
		\includegraphics[width=8.5cm, height = 3.3cm]{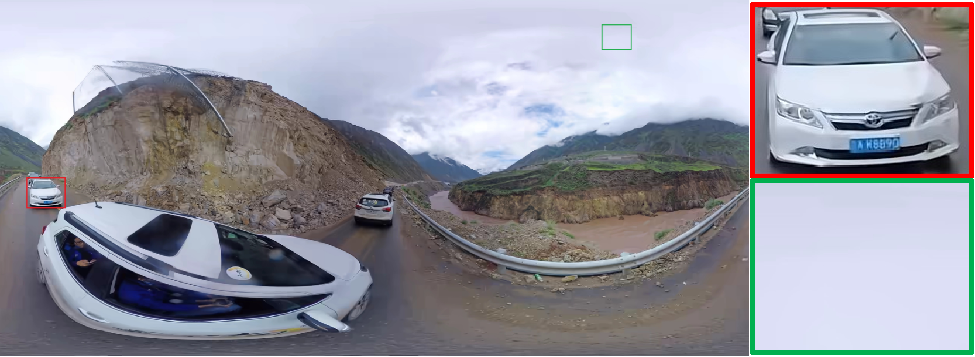}
	}
	\quad
	\subfloat[KiteFlite(VVC)]{
		\includegraphics[width=8.5cm, height = 3.3cm]{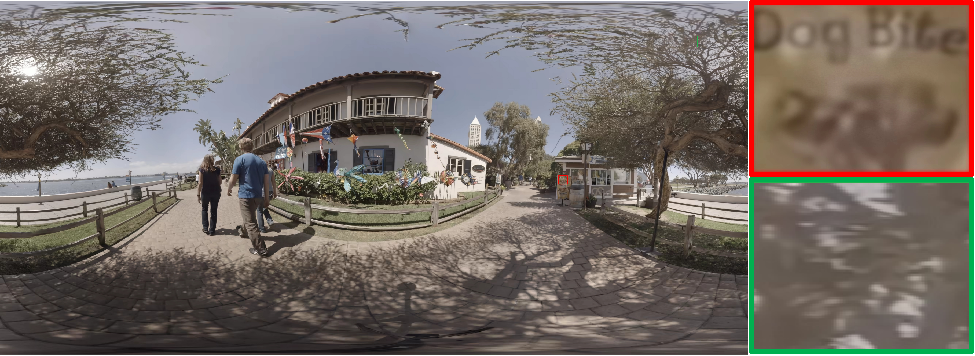}
	}
	\quad
	\subfloat[KiteFlite(SJND-VVC)]{
		\includegraphics[width=8.5cm, height = 3.3cm]{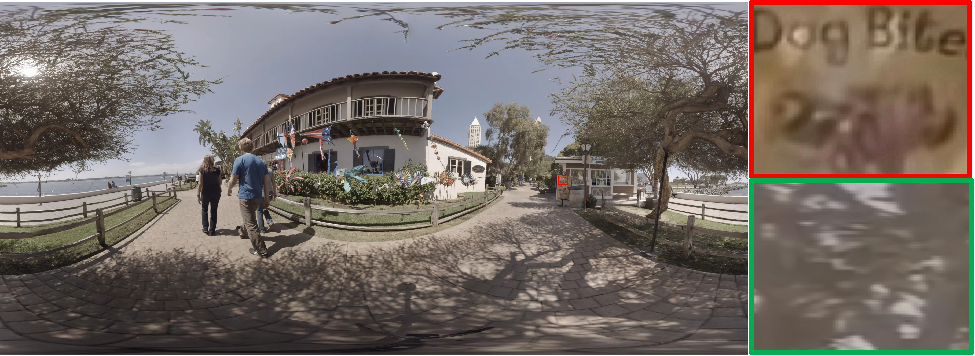}
	}
	
	\caption{Subjective Quality Comparison.}
	\label{fig16}
\end{figure*}

To better illustrate the advantages of the proposed SJND model. Fig.15 illustrates the decoded images of the DrivingInCity, DrivingInCountry, KiteFlite between the VVC and our SJND-VVC method. The text on the red bus glass is completely blurry and has a ignificant block artifact in VVC. However, the SJND-VVC can improve the block artifact that exists in the image and result in clearer edge and detail. In DrivingInCountry, we can observe that the white car outline has more severe distortion in VVC, but our SJND-VVC can significantly improves the edges and details. And since the visually sensitive areas are more refined, such as the text and graphics on billboard in KiteFlite, license plate in DrivingInCountry. In addition, for high latitude areas, SJND-VVC is able to reasonably assign QP values to ensure subjective video quality, whether CU blocks with complex textures or flat. Experimental results indicate that our SJND-VVC framework significantly reduced the bit rate with negligible loss in visual quality.

\section{Conclusion}
We propose a quantitative 2D-JND model by considering spatial contrast sensitivity, luminance adaptation, and texture masking effect. Based on the 2D-JND model, we propose a SJND model by jointly exploiting latitude projection and field of view during 360° display. The subjective test demonstrates that our SJND model is consistent with user perceptions and mitigates distortion in 360° videos. We further investigate the performance gain that the proposed SJND model can bring in video coding by embedding it into VVC. This work will benefit the future development of perceptually optimized coding for 360° videos. It can also be utilized to optimize hybrid video encoders for improved perceptual quality and perceptually-motivated video encoding schemes.

%{\appendices
%\section*{Proof of the First Zonklar Equation}
%Appendix one text goes here.
% You can choose not to have a title for an appendix if you want by leaving the argument blank
%\section*{Proof of the Second Zonklar Equation}
%Appendix two text goes here.}

\end{document}